\documentclass
[lengthestimate,tighten,
amssymb,pra,aps,twocolumn,floatfix,tightenlines,superscriptaddress]{revtex4-1}
\pdfoutput=1
\usepackage{graphics}
\usepackage{graphicx,color}
\usepackage{subfigure}
\usepackage{float}
\usepackage[ansinew]{inputenc}
\usepackage{graphicx}
\usepackage{amsmath}
\usepackage{amssymb}
\usepackage{dsfont} 
\usepackage{hyphenat}
\usepackage[percent]{overpic}
\usepackage{xcolor}
\usepackage{appendix}
\usepackage{bbold}

\newcommand{\ket}[1]{\left | #1 \right \rangle}

\begin{document}

\title{Scalable feedback control of single photon sources for photonic quantum technologies}
\author{Jacques Carolan}
\email{carolanj@mit.edu}
\affiliation{Research Laboratory of Electronics, Massachusetts Institute of Technology, Cambridge, Massachusetts 02139, USA}
\author{Uttara Chakraborty}
\affiliation{Research Laboratory of Electronics, Massachusetts Institute of Technology, Cambridge, Massachusetts 02139, USA}
\author{Nicholas C. Harris}
\affiliation{Lightmatter, 61 Chatham St 5th floor, Boston, MA 02109}
\author{Mihir Pant}
\affiliation{Research Laboratory of Electronics, Massachusetts Institute of Technology, Cambridge, Massachusetts 02139, USA}
\author{Tom Baehr-Jones}
\affiliation{Elenion Technologies, 171 Madison Avenue, Suite 1100, New York, New York 10016, USA}
\author{Michael Hochberg}
\affiliation{Elenion Technologies, 171 Madison Avenue, Suite 1100, New York, New York 10016, USA}
\author{Dirk Englund}
\affiliation{Research Laboratory of Electronics, Massachusetts Institute of Technology, Cambridge, Massachusetts 02139, USA}

\date{\today}	

\begin{abstract}
\noindent
Large-scale quantum technologies require exquisite control over many individual quantum systems. Typically, such systems are very sensitive to environmental fluctuations, and diagnosing errors via measurements causes unavoidable perturbations.
In this work we present an in situ frequency locking technique that monitors and corrects frequency variations in single photon sources based on microring resonators.
By using the same classical laser fields required for photon generation as a probe to diagnose variations in the resonator frequency, our protocol applies feedback control to correct photon frequency errors in parallel to the optical quantum computation without disturbing the physical qubit.
We implement our technique on a silicon photonic device and demonstrate sub 1~pm frequency stabilization in the presence of applied environmental noise, corresponding to a fractional frequency drift of $ <1 \%$ of a photon linewidth.
Using these methods we demonstrate feedback controlled quantum state engineering.
By distributing a single local oscillator across a single chip or network of chips, our approach enables frequency locking of many single photon sources for large-scale photonic quantum technologies. 
\end{abstract}
	
\maketitle

\section{Introduction} % (fold)
\label{sec:introduction}

Precise and robust control over individual quantum systems is a prerequisite for any scalable quantum technology.
Reducing errors in physical qubits significantly reduces the resource overhead for full-scale error correction \cite{Fowler:2012fi}, making techniques for accurate device-level calibration and control paramount.
Experimental parameters required for high-fidelity control of a quantum device may not only vary between qubits \cite{Klimov:2018jc} but also drift in time \cite{Miquel:1997di}.
Device level feedback control techniques typically measure the qubit, estimate some fidelity metric and feed back onto the control parameters to minimize the infidelity in a closed loop manner.
The success of these so called in situ control techniques \cite{Judson:1992dg, Rabitz:2000dn} hinges upon the efficiency and robustness of the fidelity estimator \cite{Ferrie:2015fe}.
While full quantum state tomography scales poorly \cite{James:2001bb, OBrien:2004cn}, techniques such as randomized benchmarking \cite{Egger:2014di, Kelly:2014fh}, direct error detection \cite{Kelly:2016fn} or efficient fidelity proxies \cite{Lu:2017ds,Li:2017bc,Dive:2017vq} have all been used to guide the system to a desired state via quantum measurement.

\begin{figure}[t!]
\includegraphics[trim=0 0 0 0, clip, width=0.9\linewidth]{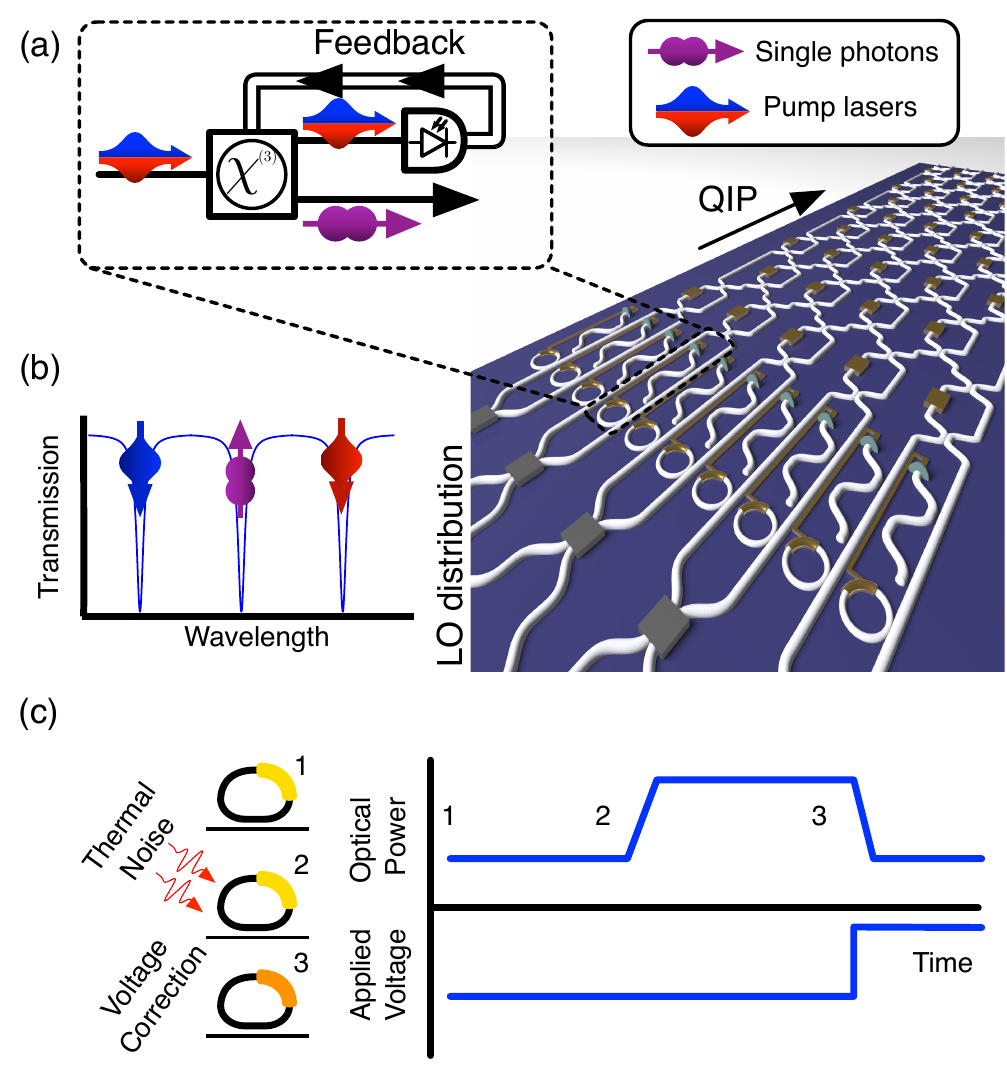}
\caption{\textbf{Proposed architecture for in situ photon source stabilization.}
(a) A pump field is coupled into a Kerr-based resonator structure, which produces correlated photons via spontaneous four-wave mixing.
The pump field is monitored via a photodiode which is fed back onto the resonator to stabilize the central frequency.  By distributing a single pump (local oscillator, LO) across an entire chip, many thousands of resonators can be frequency locked in parallel to enable large-scale quantum information processing (QIP). 
(b) Transmission spectrum of a single microring resonator.  Pump lasers are tuned to the ${i-1}^\text{th}$ and ${i+1}^\text{th}$ resonance of the ring to generate two single photons at the $i^\text{th}$ resonance.
(c) The photodiode measures an initial optical power (1), if the resonance of the MRR shifts due to, say, thermal fluctuations, the power in the pump modes increases (2), which is then corrected via a closed loop feedback on the ring phase shifter (3).
}
\label{fig:fig1}
\end{figure}

In this work we introduce a new in situ control technique for photonic quantum technologies that tracks and corrects variations in single photon sources based on microring resonators (MRRs), without the need for destructive quantum measurements.
Our protocol, shown in Fig.~\ref{fig:fig1}(a), makes use of a unique property of photonic quantum technologies where much of the error diagnosis and correction can be implemented via classical laser fields at high bandwidth, and with an intrinsically high signal-to-noise ratio. 
Using the same laser fields that seed photon generation as local oscillators to diagnose cavity fluctuations, we develop a closed loop protocol which corrects single photon frequency errors.
We implement a proof-of-concept demonstration of our technique on a silicon (Si) quantum photonic device,
and, by stabilizing on-chip cavities to sub 1~pm levels at the DC limit (corresponding to a fractional frequency drift of $<1\%$ a cavity linewidth), correct static errors between photon sources, track and correct dynamic errors and demonstrate feedback controlled quantum state engineering. 
Our corrections are performed in parallel to the quantum information processing and can be scaled to many thousands of optical components.

In photonic quantum technologies \cite{OBrien:2007ioa, Obrien:2009eu} single photons are generated via a nonlinear optical process \cite{Silverstone:2013fu, Harris:2014kj}, propagated through linear optical circuitry \cite{Laing:2010fk, Carolan:2015vga} and read out via single photon detectors \cite{Najafi:2015ey}.
Each of these core components has been demonstrated within the Si photonics platform \cite{Silverstone:2016gha} providing a plausible route towards millions of quantum optical components within a single wafer \cite{Rudolph:2017du, Sun:2015gg}.
As systems scale up \cite{Harris:2017hi, Wang:2018gh} techniques for error mitigation in quantum optical devices has become paramount.  Tools have been developed for pre-characterization of circuitry via classical laser fields  \cite{RahimiKeshari:2013bq, Grassani:2016fu} but until now, techniques for actively monitoring errors have been outstanding.

Microring resonators \cite{Bogaerts:2011eha} are a leading approach to the generation of ultra-bright \cite{Harris:2014kj, Grassani:2016fu} and pure \cite{Vernon:2017ei} single photons via the process of spontaneous four-wave mixing, with the resonance structure enabling directly engineered photon frequencies in a tens of micron-scale footprint.
In the degenerate case, where the generated photons are the same wavelength [shown in Fig.~\ref{fig:fig1}(b)], the MRR is pumped by two lasers tuned to $\omega_{p_1}, \omega_{p_2}$, corresponding to the $+n^\text{th}$ and $-n^\text{th}$ resonances of the ring.   
A photon at each frequency is spontaneously annihilated within the resonator to generate two correlated signal and idler photons at the frequency $\omega_{s,i}=(\omega_{p_1}+\omega_{p_2})/2$ in the $n=0^\text{th}$ resonance of the ring, conserving energy.
In large-scale architectures such as those required for quantum supremacy \cite{Aaronson:2011tja, Neville:2017do}, quantum simulation \cite{AspuruGuzik:2012ho, Huh:2014vk, Sparrow:2018ba} or quantum computing \cite{GimenoSegovia:2015di}, many MRRs must be tuned to precisely the same frequency.
Misalignment between resonators reduces quantum interference, which can cause errors on the photonic qubit \cite{Rohde:2012cna, Shchesnovich:2014ii}.
Moreover the efficiency and brightness of such sources scales with the quality factor of the resonator \cite{Vernon:2015fo}, placing stringent demands on the stability of MRR structures.
Fabrication variations will cause \emph{static errors} in the resonance of the MRRs, while variations in refractive index over time --- due to thermal fluctuations, the introduction of carriers, electrical noise or cross-talk between devices --- will introduce \emph{dynamic errors}.
% section introduction (end)

Our approach shown in Fig.~\ref{fig:fig1}(c) monitors the pump frequency modes with a low-loss drop filter and photodiode.
If the central frequency of the resonator shifts the optical power on the photodiode will increase, and an electrical signal is fed back onto the phase shifter in a closed loop manner to decrease the optical power.
This minimization can be implemented in either software (e.g. computational optimization) or hardware (e.g. lock-in amplifier \cite{Padmaraju:2013ba}).
Our closed loop protocol scales with a time complexity $\mathcal{O}(1)$ in the number of MRRs, and is typically bandwidth-limited by the control phase modulator.
Each constituent component has already been demonstrated in standard CMOS Si photonic processes: low-loss filtering \cite{Horst:2013il}, fast photodiodes \cite{Michel:2010ib} and phase modulation [including thermo-optic (kHz \cite{Harris:2014kz}), microelectromechanical (MHz \cite{Seok:2016he}) and carrier-based (GHz \cite{Thomson:2011hk})].
Moreover, the classical probe signal provides an intrinsically high signal-to-noise ratio compared with direct detection of the photons.
% section the_protocol (end)

\section{The Device} % (fold)
\begin{figure}[t!]
\includegraphics[trim=0 0 0 0, clip, width=0.9\linewidth]{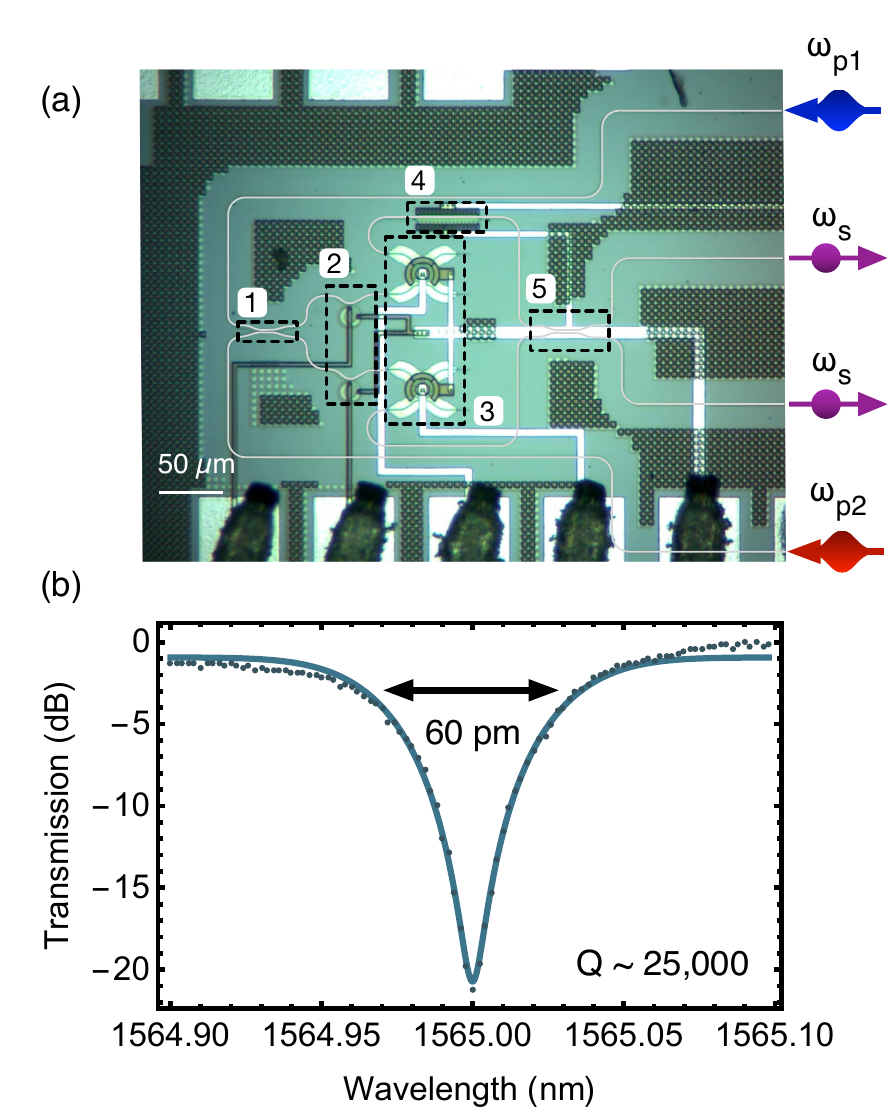}
\caption{\textbf{Quantum state engineering photonic device.}
(a) An optical micrograph of the silicon photonic device which incorporates five thermo-optically controlled phases shifters and four microring resonators (two for photon generation and two for pump suppression) in just $0.08~\text{mm}^2$.  
Marked components represent the five stages required for quantum state engineering: (1) Pump mixing on a directional coupler, (2) photon generation in two MRRs, (3) partial pump suppression in two further MRRs, (4) differential phase shift and (5) final directional coupler for quantum interference.
(b) An optical spectrograph of the two generation rings aligned to $1565$~nm alongside expected fit.
}
\label{fig:fig2}
\end{figure}

\label{sec:the_device}
For our proof-of-concept demonstration we use a quantum state engineering Si photonic device, alongside off-chip pump separation and monitoring.
The device produces correlated pairs of photons via the inverse Hong-Ou-Mandel effect \cite{Silverstone:2013fu} and comprises five stages as shown in Fig.~\ref{fig:fig2}(a). 
The first mixes the two pumps on a 50/50 directional coupler.  
Next, the mixed pumps impinge on a photon generation MRR in each arm of a Mach-Zehnder interferometer.
The pump power is partially reduced via demux filters to prevent further photon generation in the waveguides, yet remains at a level sufficient to be monitored via off-chip photodiodes.
The state passes through a differential phase $\phi$, and by operating in the weak pumping regime such that an appreciable probability exists only of producing two photons, the quantum state after the two rings is $\ket{\psi}_\text{ring}=(\ket{20}_{1,2}+e^{2i\phi}\ket{02}_{1,2})/\sqrt{2}$, where $\ket{n}_m$ represents $n$ photons in the $m^\text{th}$ optical mode.
Finally, the state is incident on a 50/50 directional coupler which yields the state 
\begin{equation}
	\label{eq:hom}
	\ket{\psi(\phi)}_\text{out}=\cos{\phi} (\ket{20}-\ket{02})/\sqrt{2} + \sin{\phi}\ket{11}.
\end{equation}
Control of the differential phase therefore enables state engineering, including tuning between path entangled states ($\phi=0$) and separable states ($\phi=\pi$).

\vspace{-0.0cm}

The chip, fabricated in a standard CMOS Si photonics process, contains four MRRs and five thermo-optic phase shifters all within $0.08~\text{mm}^2$ [see Fig.~\ref{fig:fig2}(a)].
The spectrum of the photon generation MRR is shown in Fig.~\ref{fig:fig2}(b).
Each ring has with a linewidth $\Delta \lambda = 60~\text{pm}$, yielding a quality factor of $Q\approx 2.5\times 10^4$.
Light is in/out-coupled via a custom-built silicon nitride optical interposer, which matches both the mode field diameter and pitch of the Si waveguides to give a loss of $-2.5\pm0.5~\text{dB}$ per facet (error determined by multiple measurements).
At the input two tunable telecommunication lasers are pre-filtered to reduce optical sidebands at the photon generation wavelength.
At the output photons are first filtered to enable pump monitoring and reduce background, then coupled into superconducting nanowire single photon detectors with $\sim 75\%$ quantum efficiency.
See Appendix for further experimental details.
% section the_device (end)

\begin{figure*}[t!]
\includegraphics[trim=0 0 0 0, clip, width=0.9\linewidth]{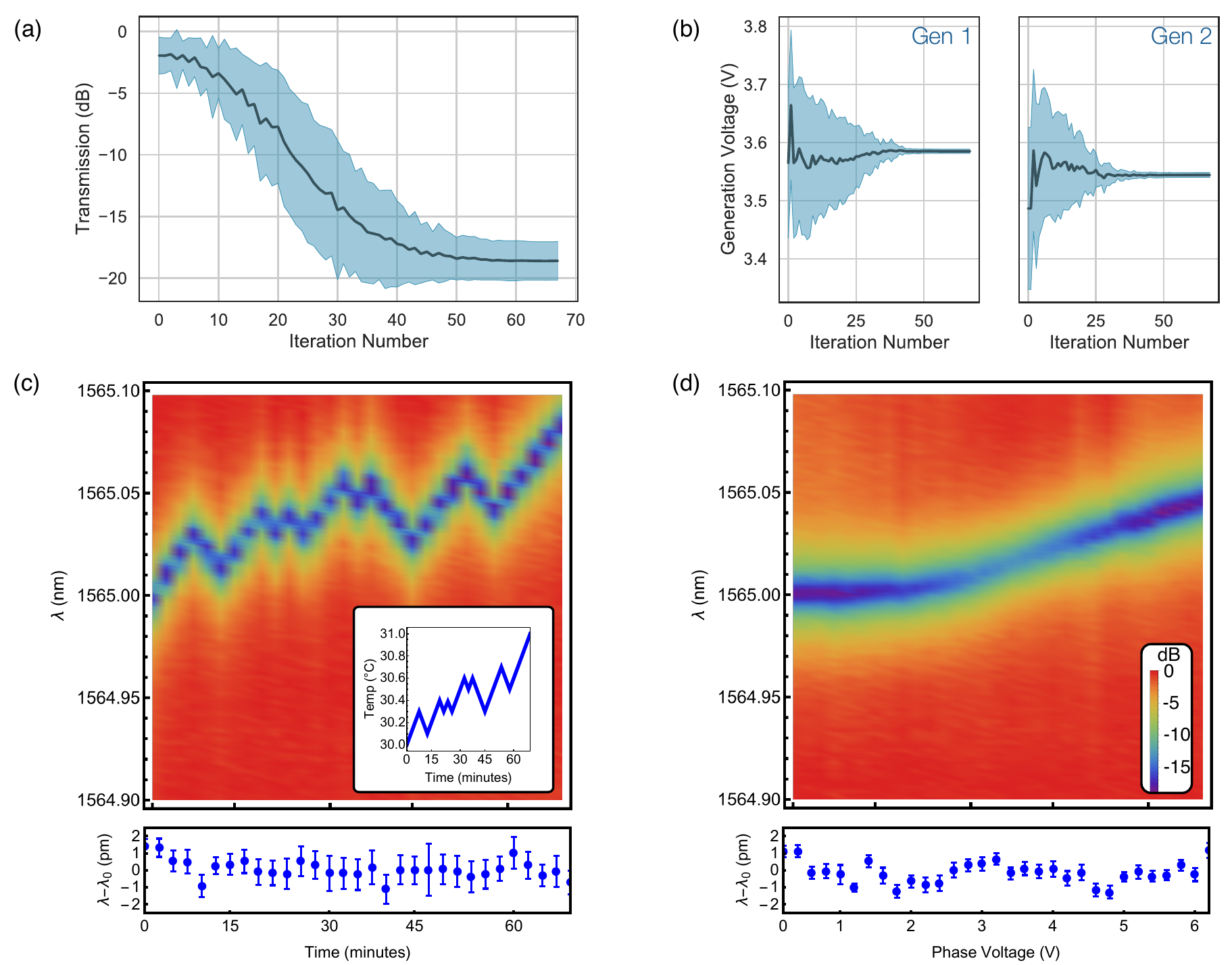}
\caption{\textbf{Static and dynamic feedback correction.}
(a) The mean of 66 instances of static frequency feedback correction, with initial guess voltages for each run randomly and independently chosen (see text).
The shaded region represents $\pm 1 \sigma$.  
With the pump laser set to the desired alignment frequency of $\lambda_0 = 1565$~nm, the voltage on each generation MRR is optimized to minimize the sum of the optical power in two output modes.
(b) The mean change in voltages for each generation MRR during all 66 alignment protocols.
Solution voltages not only vary between MRRs (a static offset due to fabrication variations) but also over the course of the experiment due to a systematic change in laboratory conditions. 
(c) A spectrograph of the MRRs as a function of applied thermal noise (inset) over the course of one hour in the absence of dynamic stabilization.
Given the same applied noise model, the below plot shows the variation in central resonance when dynamic frequency stabilization is applied.  Error bars are given by the error in the resonance fit.
(d) A spectrograph of the MRRs as a voltage is applied to an adjacent thermo-optic phase-shifter.   Thermal cross-talk causes the resonance of the MRRs to shift, which should otherwise remain untouched by the phase shifter.
The below plot shows the variation when dynamic frequency stabilization is applied.
In each instance the dynamic stabilization gives a two orders of magnitude increase in the resonance stability.
}
\label{fig:fig3}
\end{figure*}

\section{The Protocol} % (fold)
\label{sec:the_protocol}

As a first test of our frequency locking protocol we correct static errors in the resonance position of the generation rings which can occur due fabrication variations such as waveguide surface-roughness \cite{Little:1997kk}.
In principle, accurate characterization of wavelength-voltage tuning curves can correct for this effect, but as we show, noise sources such as thermal crosstalk and electrical noise will complicate this process, necessitating an in situ approach.
For this test the feedback correction protocol is run 100 times. 
Each run sets the pump laser to the desired generation wavelength, and initial voltages for the two generation rings are chosen randomly from normal distributions centered on 3.60~V and 3.56~V respectively with a standard deviation of 0.2~V. 
Computational optimization is used to iteratively arrive at the generation ring voltage combination that minimizes the sum of the optical output powers of the MRRs as measured by an off-chip photodiode array.
The gradient-free Nelder-Mead algorithm was empirically determined to converge quickly and be robust in the presence of experimental noise. 
As shown in Fig.~\ref{fig:fig3}(a), out of the 100 attempted runs 62 succeed, requiring an average of 57 iterations to converge.
Figure~\ref{fig:fig3}(b) tracks the voltages of each generation MRR during optimization.
The final voltage of each ring differs by 40~mV, demonstrating the importance of static error correction.  
Moreover, repeatedly running this protocol over the course of 7 hours, we observe a total reduction in the voltages by 18~mV, likely due to a systematic drift in laboratory temperature.

In Fig.~\ref{fig:fig3}(c,d) we simulate two classes of dynamic error typically seen in photonic quantum systems: (1) environmental temperature fluctuations and (2) crosstalk between thermo-optic phase shifters. 
We induce temperature fluctuations by varying the chip temperature through an auxiliary Peltier control system onto which the device is mounted.
In increments and decrements of $0.1^{\circ}$C, we program a random walk in temperature over the course of one hour for a net increase of $1^{\circ}$C.
One instance of this random walk is shown in the Fig~\ref{fig:fig3}(c) inset.
Figure~\ref{fig:fig3}~(c) plots spectrographs for this instance which show the shift in the central resonance of the MRRs as a result of this temperature variation in the absence of dynamic frequency stabilization and in the presence of our in situ approach.  
The implementation of our protocol leads to a standard deviation in the central resonance wavelength of $0.56~\text{pm}$ ($9.4\times 10^{-3} \Delta \lambda$), compared to a total variation of $84.0~\text{pm}$ ($1.4 \Delta \lambda $) in the absence of any correction protocol. This corresponds to a two-orders of magnitude increase in resonance stability.

Similarly, we induce thermal crosstalk by sweeping the phase shifter voltage from 0 to 6.5 V.  Figure 3 shows the central wavelength shift in (d) the uncorrected case, and (d) the in-situ corrected case.  Dynamic frequency stabilization yields a stability of $0.65~\text{pm}$ ($1.1\times 10^{-2} \Delta \lambda$), a 70-fold improvement compared with a total variation of $45~\text{pm}$ ($0.75\Delta \lambda$) in the uncorrected case. 

We contrast the performance of our in situ correction technique with the results obtained using pre-determined tuning curve models (see Appendix for details) to align the rings, with the same temperature or phase shifter voltage adjustment. After each adjustment the generation ring voltages are set to the values according to the pre-determined functions. While alignment using pre-determined functions leads to a 15-fold and 5-fold improvement over the uncorrected case for the temperature and voltage error respectively, our iterative protocol still outperforms the tuning curve-based correction by an order of magnitude in both instances.
Moreover, our technique can naturally be applied to dynamic corrections where no noise model is known.

\begin{figure}[t!]
\includegraphics[trim=0 0 0 0, clip, width=1.0\linewidth]{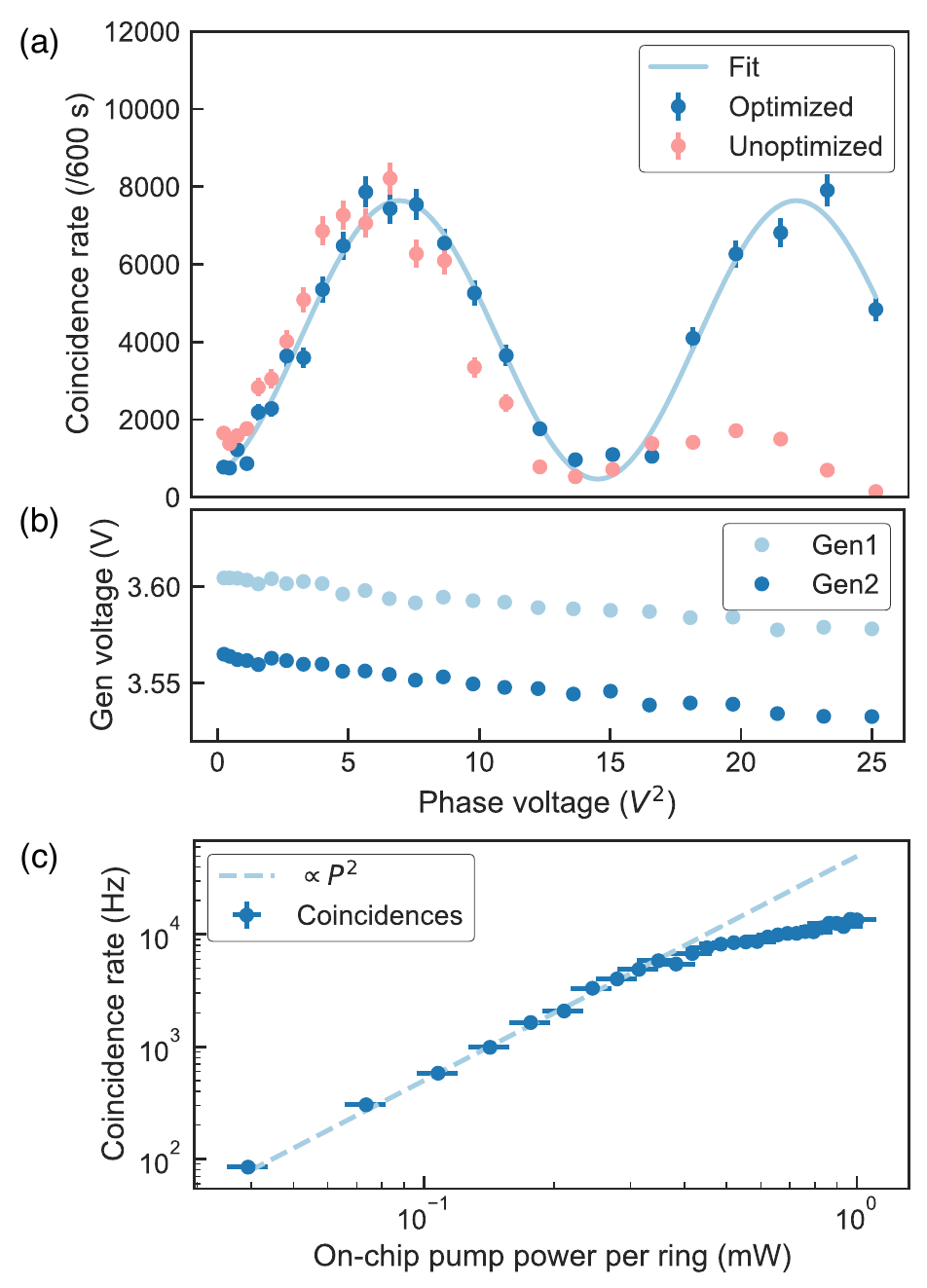}
\caption{\textbf{Quantum state engineering.}
(a) Coincidence count rate plotted as a function of the square of the differential phase voltage, with (blue) and without (red) frequency stabilization, alongside a sinusoidal fit (light blue).  
Coincidences have been normalized for detector channel inefficiencies and error bars assume Poissonian counting statistics.
The symmetry in the locked fringe can clearly be observed in comparison to the unlocked.
(b) Variation in MRR control voltages over the course of the differential phase sweep when frequency locking is applied.
(c) Coincidence count rate plotted as a function of input power per ring (blue points) and an expected quadratic dependency based on a purely four-wave mixing process (light blue line).
}
\label{fig:fig4}
\end{figure}

The merit of the in situ approach is that it can be performed in parallel to the quantum computation. 
To demonstrate this our protocol is applied to the task of quantum state engineering.  
According to Eq.~\eqref{eq:hom}, a linear variation in the differential phase $\phi$ causes a sinusoidal change in the probability amplitude of the $\ket{11}$ state, and a sin-squared change in the coincidence probability. 
Control of the thermo-optic phase shifter thus provides a direct means to engineer the photonic quantum state.
In the absence of frequency control (Fig.~\ref{fig:fig4}(a), red) thermal cross-talk from the differential phase decouples the MRRs and causes an asymmetry in the interference fringe.
To quantify this effect we introduce the asymmetric contrast $C_\text{asy}=|C_1-C_2|/\text{max}(C_1,C_2)$,  which is the normalized difference between the coincidence counts $C_1$ at $\phi=\pi/2$ and counts $C_2$ at $\phi=3 \pi/2$, where $C_\text{asy}=0$ in the ideal case.
In the absence of correction $C_\text{asy}=0.791$.

The frequency control protocol is implemented at each step of the phase sweep (Fig.~\ref{fig:fig4}(a), blue) which corrects the generation voltages [Fig.~\ref{fig:fig4}(b)] and recovers the symmetry of the interference fringe, yielding a contrast $C_\text{asy}=5.61\times 10^{-3}$.
The quantum visibility quantifies the indistinguishability of the photons and is given by $V_\text{q}=(C_\text{max}-C_\text{min})/C_\text{max}$ where $C_\text{max} (C_\text{min})$ is the maximum (minimum) measured coincidence counts.
The interference fringe is fitted (Fig.~\ref{fig:fig4}(a), blue line) to account for the nonlinear phase-voltage relation of the thermo-optic phase shifter \cite{Harris:2014kz}, and the quantum visibility is extracted as $V_\text{q}=0.938\pm 0.021$.
The deviation from unity visibility is primarily due to higher order photon events, which occur due to the high pump power required to obtain a reasonable signal-to-noise ratio in the presence of lossy off-chip filters.
In future, the monolithic integration of lasers \cite{Zhou:2015gj}, single photon detectors \cite{Najafi:2015ey} and filters \cite{Harris:2014kj, Piekarek:2017ch} will significantly reduce optical power constraints.

Finally in Fig.~\ref{fig:fig4}(c), with $\phi=\pi/2$, we measure the coincidence count rate as a function of the input pump power.  
At each optical power setting we apply the frequency stabilization protocol to account for the refractive index change in the MRRs due to a combination of Kerr, thermal and free-carrier dispersion effects  \cite{Wang:2016jma}.
We reach an off-chip photon generation rate of 13.5 kHz (corrected for detector channel inefficiencies) which is primarily limited by two photon absorption.
This can be seen in Fig.~\ref{fig:fig4}(c) where we plot the measured coincidence count rate against the expected quadratic dependence (Fig.~\ref{fig:fig4}(c), blue dashed) observing deviations at powers greater than 200~$\mu$W.
Significant progress is being made on mid-IR silicon photonics, that will mitigate the effect of two photon absorption which becomes negligible at wavelengths longer than 2.2~$\mu$m \cite{Lin:2017io, Zou:2018ke}.

\section{Conclusion} % (fold)
\label{sec:conclusion}
We have proposed and demonstrated an in situ control technique for photonic quantum technologies that
uses the same classical laser fields required for photon generation as a probe to track, diagnose and correct frequency variations in single photon sources.
While feedback control in our device is applied off-chip, in situ feedback was recently demonstrated in an integrated CMOS photonics platform \cite{Sun:2015gg}. 
Electronic control circuitry either integrated on-chip \cite{Atabaki:2018jf} or via flip-chip approaches \cite{Carroll:2016fa}, would therefore allow large numbers of heralded single photon sources to be frequency locked to a common local oscillator.
The combination of Kerr nonlinear optics in silicon rings with CMOS logic and single photon detection \cite{Schuck:2013be, Schelew:2015jw, Najafi:2015ey, DiZhu:2018ja}, could enable on-demand high fidelity single photon sources based on multiplexed spontaneous four-wave mixing \cite{Heuck:2018km},
which form the basis of proposed all-optical quantum computing \cite{Rudolph:2017du} and quantum repeater architectures \cite{Pant:2017ca}.

% section conclusion (end)

\begin{acknowledgments}
This work was supported by the AFOSR MURI for Optimal Measurements for Scalable Quantum Technologies (FA9550-14-1-0052) and by the AFOSR program FA9550-16-1-0391, supervised by Gernot Pomrenke. J.C. is supported by EU H2020 Marie Sklodowska-Curie grant number 751016. U.C. is supported by the National Defense Science and Engineering Graduate Fellowship.
We gratefully acknowledge A. Pyke at Aerospace Semiconductor, C. Panuski, C. Chen, F. Wong and J. Combes. 
\end{acknowledgments}

\clearpage

\appendix
\section{Device Details} % (fold)
 \label{sec:appendix}
The device, shown in Fig.~\ref{fig:fig1}(a), is fabricated in a standard CMOS silicon photonics process, and consists of two microring resonators for photon generation with radius $R=11~\mu$m, coupled to a 500~nm wide $\times$ 220~nm silicon bus waveguide.
Each ring has a Q factor of $2.5\times10^4$, and a free spectral range $\text{FSR}=8.8$~nm.
After just $40~\mu m$ the bus waveguide is coupled to a demultiplexing ring ($R=8~\mu$m, FSR$=12$~nm) to separate single photons and pump light, and photons via the drop port are routed to a phase shifter and directional coupler for state engineering.
All four rings are thermo-optically controlled by embedded resistive heaters formed by doped silicon regions contacting the metal interconnect layer. To minimize losses due to free-carrier absorption, a low dopant concentration in the waveguide region overlapping with the optical mode is employed.
The combination of both generation and demultiplexing rings enables a pump suppression of 37~dB, mitigating further incoherent photon generation within the bus waveguide.
The experimental setup consists of two tuneable telecom lasers set to $\lambda_{p_1}=1582.3$~nm and $\lambda_{p_2}=1547.7$~nm, at $+2$ and $-2$ FSR of the tuned generation rings, for degenerate pair photon generation at $\lambda_{s,i}=1565.0$~nm.
% %
 Pump lasers are passed through tuneable band pass filters which provides a total of $100$~dB suppression of unwanted sidebands occurring due to amplified spontaneous emission.

Laser light is edge coupled into the chip via custom built SiN interposers, which reduces the optical mode field diameter to better match the on-chip tapered mode convertor, achieving an estimated loss per facet of $-2.5\pm 0.5$~dB.
% %
 The device is mounted on top of a Peltier cooling unit to maintain thermal stability, and the thermo-optic phase shifters are controlled by a custom-built multi-channel digital to analogue converter which provides 16-bits voltage precision.
% %
 Both correlated photons and pump light are out-coupled and passed through narrow linewidth filters, which along with on-chip filtering, provides a total pump suppression of $\sim100$~dB.
% %
 Photons are sent to two superconducting nanowire single photon detectors with quantum efficiencies of $\eta=75~\%$, and the signals are time-tagged using a time-correlated single photon counting module.
% % section appendix (end)

 \begin{figure}[t]
 \includegraphics[trim=0 0 0 0, clip, width=1.0\linewidth]{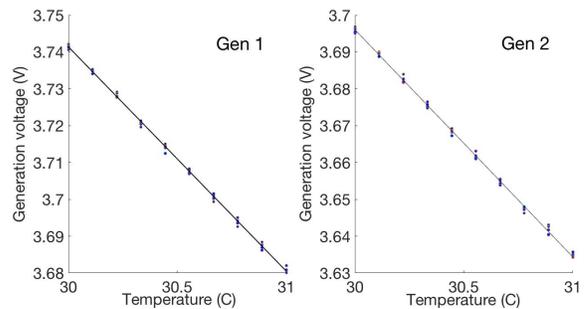}
 \caption{\textbf{Temperature tuning.} Five sets of temperature-voltage data points for each generation ring alongside a linear fit.
 }
 \label{fig:fig5}
 \end{figure}

 \section{Theoretical Coupled Ring Model} % (fold)
 \label{sec:model}

% % section model (end)
%
By modeling the transmission of coupled microring resonators, we can show that there is one and only one possible generation ring voltage combination that leads to a minimum in the rings' combined transmitted power, and hence there are no local minima that the Nelder-Mead search algorithm could potentially converge to. The transmission function of a single ring can be taken to be a Lorentzian:
\begin{equation}
T (\lambda) = \frac{-0.5 \Gamma}{(\lambda - \lambda_{las})^{2} + (0.5\Gamma)^{2}}  
 \end{equation} where $\Gamma$ and $\lambda_{las}$ are the width parameter and laser wavelength respectively.
 The dependence of the rings' central wavelengths ($\lambda_{1}$ and $\lambda_{2}$)  on ring voltages can be modelled as
\begin{eqnarray}
\lambda_{1} & = & \lambda_{01} + \gamma_{1} V_{1}^{2} + \alpha_{12} V_{2}^{2}  \\  
\lambda_{2} & = & \lambda_{02} + \gamma_{2} V_{2}^{2} + \alpha_{12} V_{1}^{2}
\end{eqnarray}
where $\lambda_{01}$ and $\lambda_{02}$ are the central resonances of the rings with no applied voltage tuning, coefficients $\gamma_{1}$ and $\gamma_{2}$ correspond to the strength of the rings' wavelength dependence on voltage applied to themselves, and the coefficient $\alpha_{12}$ corresponds to the strength of the each ring's wavelength dependence on voltage applied to the other. The voltage-squared dependence of the central wavelength on voltage arises from linearity of the wavelength shift with temperature, and hence with the dissipated power. In a physically realistic case, both the ratios $\frac{\gamma_{1}}{\alpha_{12}}$ and $\frac{\gamma_{2}}{\alpha_{12}}$ will be much greater than both $\frac{\lambda_{las}-\lambda_{01}}{\lambda_{las}-\lambda_{02}}$  and $\frac{\lambda_{las}-\lambda_{02}}{\lambda_{las}-\lambda_{01}}$. 
The total transmission of two rings in series is given as:
\begin{equation}
T (\lambda_{1}, \lambda_{2}) = \frac{-0.5 \Gamma}{(\lambda_{1} - \lambda_{las})^{2} + (0.5\Gamma)^{2}} * \frac{-0.5 \Gamma}{(\lambda_{2} - \lambda_{las})^{2} + (0.5\Gamma)^{2}} 
 \end{equation} and the total transmission in parallel as:
 \begin{equation}
T (\lambda_{1}, \lambda_{2}) = \frac{-0.5 \Gamma}{(\lambda_{1} - \lambda_{las})^{2} + (0.5\Gamma)^{2}} + \frac{-0.5 \Gamma}{(\lambda_{2} - \lambda_{las})^{2} + (0.5\Gamma)^{2}}. 
 \end{equation}
Both the series and parallel transmission functions have critical points where the conditions $\frac{\partial T}{\partial \lambda_{1}}=0$ and $\frac{\partial T}{\partial \lambda_{2}}=0$ hold. In order to satisfy both conditions, we require $V_{1}=0$ or $\lambda_{1}=\lambda_{01} + \gamma_{1} V_{1}^{2} + \alpha_{12} V_{2}^{2}=\lambda_{las}$, and $V_{2}=0$ or $\lambda_{1}=\lambda_{01} + \gamma_{1} V_{1}^{2} + \alpha_{12} V_{2}^{2}=\lambda_{las}$. Out of the four possible combinations, only one gives a minimum (the others are a maximum and saddle points): 
\begin{eqnarray}
\lambda_{1} & = & \lambda_{01} + \gamma_{1} V_{1}^{2} + \alpha_{12} V_{2}^{2}=\lambda_{las}  \\
\lambda_{2} & = & \lambda_{02} + \gamma_{2} V_{2}^{2} + \alpha_{12} V_{1}^{2}=\lambda_{las}
\end{eqnarray} 
Given the physically realistic stipulations on $\gamma_{1}$, $\gamma_{2}$, $\alpha_{12}$, $\lambda_{las}-\lambda_{01}$ and $\lambda_{las}-\lambda_{02}$, the two equations above are guaranteed to have a solution with non-zero values of $V_{1}$ and $V_{2}$, which corresponds to tuning both rings to the laser wavelength.
Hence, there is only one global minimum value of the transmission function for non-negative voltages, and no local minima. This guarantees that if our search converges, it will have converged to the true global minimum. 
This model may be generalised to an arbitrary number of ring resonators in series or parallel, such that the total transmission of N rings in series will be given by 
  \begin{equation}
 T(\lambda_{1}, \lambda_{2}, \cdots, \lambda_{N}) = \prod_{i = 1}^{N} \frac{-0.5 \Gamma}{(\lambda_{i} - \lambda_{las})^{2} + (0.5\Gamma)^{2}} 
 \end{equation} and the total transmission in parallel by 
  \begin{equation}
 T(\lambda_{1}, \lambda_{2}, \cdots, \lambda_{N}) = \sum_{i = 1}^{N} \frac{-0.5 \Gamma}{(\lambda_{i} - \lambda_{las})^{2} + (0.5\Gamma)^{2}} 
 \end{equation} {
As in the two-ring case above, the sole minimum of the transmission function is achieved when all rings are individually tuned to the laser wavelength, and there are no local minima.
 
\begin{figure}[t!]
\includegraphics[trim=0 0 0 0, clip, width=1.0\linewidth]{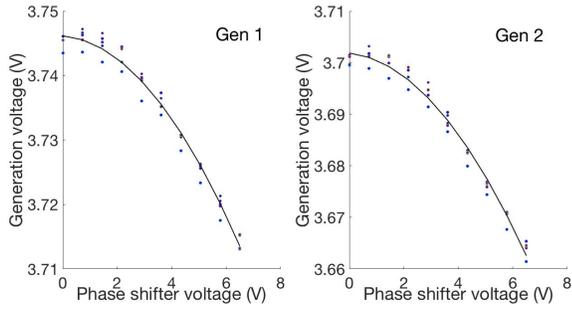}
\caption{\textbf{Phase voltage tuning.} Five sets of Phase shifter-voltage data points for each generation ring alongside a quadratic fit.
}
\label{fig:fig6}
\end{figure}
 
 \section{Tuning Curves}
 \label{sec:tuning}
 By sweeping the temperature, $T$, from 30 to 31 degrees and iteratively aligning the rings at each voltage using the Nelder-Mead algorithm, we obtain generation ring voltage ($V_{1}$ and $V_{2}$) versus temperature data (Fig~\ref{fig:fig5}). Based on 5 such sweeps, we obtain the following best-fit linear model for the dependence of the ring voltages on temperature:\begin{eqnarray}
 V_{1}(T) = -0.06090 T + 5.568 \\
 V_{2}(T) = -0.06166 T + 5.546
 \end{eqnarray}

 Similarly, by sweeping the phase shifter voltage, $V_{p}$, from 0 to 6.5 volts and aligning the rings at each voltage using the Nelder-Mead algorithm, we obtain ring voltage versus phase shifter voltage data (Fig~\ref{fig:fig6}). Based on 5 sweeps, we obtain a best-fit quadratic model for the dependence of the ring voltages on the phase shifter voltage:
 \begin{eqnarray}
V_{1}(V_{p}) = -0.0007192 V_{p}^{2} - 0.0003439 V_{p} + 3.746  \\
V_{2}(V_{p}) = -0.0008414 V_{p}^{2} - 0.000576 V_{p} + 3.702
 \end{eqnarray}


\begin{thebibliography}{58}%
\makeatletter
\providecommand \@ifxundefined [1]{%
 \@ifx{#1\undefined}
}%
\providecommand \@ifnum [1]{%
 \ifnum #1\expandafter \@firstoftwo
 \else \expandafter \@secondoftwo
 \fi
}%
\providecommand \@ifx [1]{%
 \ifx #1\expandafter \@firstoftwo
 \else \expandafter \@secondoftwo
 \fi
}%
\providecommand \natexlab [1]{#1}%
\providecommand \enquote  [1]{``#1''}%
\providecommand \bibnamefont  [1]{#1}%
\providecommand \bibfnamefont [1]{#1}%
\providecommand \citenamefont [1]{#1}%
\providecommand \href@noop [0]{\@secondoftwo}%
\providecommand \href [0]{\begingroup \@sanitize@url \@href}%
\providecommand \@href[1]{\@@startlink{#1}\@@href}%
\providecommand \@@href[1]{\endgroup#1\@@endlink}%
\providecommand \@sanitize@url [0]{\catcode `\\12\catcode `\$12\catcode
  `\&12\catcode `\#12\catcode `\^12\catcode `\_12\catcode `\%12\relax}%
\providecommand \@@startlink[1]{}%
\providecommand \@@endlink[0]{}%
\providecommand \url  [0]{\begingroup\@sanitize@url \@url }%
\providecommand \@url [1]{\endgroup\@href {#1}{\urlprefix }}%
\providecommand \urlprefix  [0]{URL }%
\providecommand \Eprint [0]{\href }%
\providecommand \doibase [0]{http://dx.doi.org/}%
\providecommand \selectlanguage [0]{\@gobble}%
\providecommand \bibinfo  [0]{\@secondoftwo}%
\providecommand \bibfield  [0]{\@secondoftwo}%
\providecommand \translation [1]{[#1]}%
\providecommand \BibitemOpen [0]{}%
\providecommand \bibitemStop [0]{}%
\providecommand \bibitemNoStop [0]{.\EOS\space}%
\providecommand \EOS [0]{\spacefactor3000\relax}%
\providecommand \BibitemShut  [1]{\csname bibitem#1\endcsname}%
\let\auto@bib@innerbib\@empty
%</preamble>
\bibitem [{\citenamefont {Fowler}\ \emph {et~al.}(2012)\citenamefont {Fowler},
  \citenamefont {Mariantoni}, \citenamefont {Martinis},\ and\ \citenamefont
  {Cleland}}]{Fowler:2012fi}%
  \BibitemOpen
  \bibfield  {author} {\bibinfo {author} {\bibfnamefont {A.~G.}\ \bibnamefont
  {Fowler}}, \bibinfo {author} {\bibfnamefont {M.}~\bibnamefont {Mariantoni}},
  \bibinfo {author} {\bibfnamefont {J.~M.}\ \bibnamefont {Martinis}}, \ and\
  \bibinfo {author} {\bibfnamefont {A.~N.}\ \bibnamefont {Cleland}},\
  }\href@noop {} {\bibfield  {journal} {\bibinfo  {journal} {Phys. Rev. A}\
  }\textbf {\bibinfo {volume} {86}},\ \bibinfo {pages} {032324} (\bibinfo
  {year} {2012})}\BibitemShut {NoStop}%
\bibitem [{\citenamefont {Klimov}\ \emph {et~al.}(2018)\citenamefont {Klimov},
  \citenamefont {Kelly}, \citenamefont {Chen}, \citenamefont {Neeley},
  \citenamefont {Megrant}, \citenamefont {Burkett}, \citenamefont {Barends},
  \citenamefont {Arya}, \citenamefont {Chiaro}, \citenamefont {Chen},
  \citenamefont {Dunsworth}, \citenamefont {Fowler}, \citenamefont {Foxen},
  \citenamefont {Gidney}, \citenamefont {Giustina}, \citenamefont {Graff},
  \citenamefont {Huang}, \citenamefont {Jeffrey}, \citenamefont {Lucero},
  \citenamefont {Mutus}, \citenamefont {Naaman}, \citenamefont {Neill},
  \citenamefont {Quintana}, \citenamefont {Roushan}, \citenamefont {Sank},
  \citenamefont {Vainsencher}, \citenamefont {Wenner}, \citenamefont {White},
  \citenamefont {Boixo}, \citenamefont {Babbush}, \citenamefont {Smelyanskiy},
  \citenamefont {Neven},\ and\ \citenamefont {Martinis}}]{Klimov:2018jc}%
  \BibitemOpen
  \bibfield  {author} {\bibinfo {author} {\bibfnamefont {P.~V.}\ \bibnamefont
  {Klimov}}, \bibinfo {author} {\bibfnamefont {J.}~\bibnamefont {Kelly}},
  \bibinfo {author} {\bibfnamefont {Z.}~\bibnamefont {Chen}}, \bibinfo {author}
  {\bibfnamefont {M.}~\bibnamefont {Neeley}}, \bibinfo {author} {\bibfnamefont
  {A.}~\bibnamefont {Megrant}}, \bibinfo {author} {\bibfnamefont
  {B.}~\bibnamefont {Burkett}}, \bibinfo {author} {\bibfnamefont
  {R.}~\bibnamefont {Barends}}, \bibinfo {author} {\bibfnamefont
  {K.}~\bibnamefont {Arya}}, \bibinfo {author} {\bibfnamefont {B.}~\bibnamefont
  {Chiaro}}, \bibinfo {author} {\bibfnamefont {Y.}~\bibnamefont {Chen}},
  \bibinfo {author} {\bibfnamefont {A.}~\bibnamefont {Dunsworth}}, \bibinfo
  {author} {\bibfnamefont {A.}~\bibnamefont {Fowler}}, \bibinfo {author}
  {\bibfnamefont {B.}~\bibnamefont {Foxen}}, \bibinfo {author} {\bibfnamefont
  {C.}~\bibnamefont {Gidney}}, \bibinfo {author} {\bibfnamefont
  {M.}~\bibnamefont {Giustina}}, \bibinfo {author} {\bibfnamefont
  {R.}~\bibnamefont {Graff}}, \bibinfo {author} {\bibfnamefont
  {T.}~\bibnamefont {Huang}}, \bibinfo {author} {\bibfnamefont
  {E.}~\bibnamefont {Jeffrey}}, \bibinfo {author} {\bibfnamefont
  {E.}~\bibnamefont {Lucero}}, \bibinfo {author} {\bibfnamefont {J.~Y.}\
  \bibnamefont {Mutus}}, \bibinfo {author} {\bibfnamefont {O.}~\bibnamefont
  {Naaman}}, \bibinfo {author} {\bibfnamefont {C.}~\bibnamefont {Neill}},
  \bibinfo {author} {\bibfnamefont {C.}~\bibnamefont {Quintana}}, \bibinfo
  {author} {\bibfnamefont {P.}~\bibnamefont {Roushan}}, \bibinfo {author}
  {\bibfnamefont {D.}~\bibnamefont {Sank}}, \bibinfo {author} {\bibfnamefont
  {A.}~\bibnamefont {Vainsencher}}, \bibinfo {author} {\bibfnamefont
  {J.}~\bibnamefont {Wenner}}, \bibinfo {author} {\bibfnamefont {T.~C.}\
  \bibnamefont {White}}, \bibinfo {author} {\bibfnamefont {S.}~\bibnamefont
  {Boixo}}, \bibinfo {author} {\bibfnamefont {R.}~\bibnamefont {Babbush}},
  \bibinfo {author} {\bibfnamefont {V.~N.}\ \bibnamefont {Smelyanskiy}},
  \bibinfo {author} {\bibfnamefont {H.}~\bibnamefont {Neven}}, \ and\ \bibinfo
  {author} {\bibfnamefont {J.~M.}\ \bibnamefont {Martinis}},\ }\href@noop {}
  {\bibfield  {journal} {\bibinfo  {journal} {Phys. Rev. Lett.}\ }\textbf
  {\bibinfo {volume} {121}},\ \bibinfo {pages} {090502} (\bibinfo {year}
  {2018})}\BibitemShut {NoStop}%
\bibitem [{\citenamefont {Miquel}\ \emph {et~al.}(1997)\citenamefont {Miquel},
  \citenamefont {Paz},\ and\ \citenamefont {Zurek}}]{Miquel:1997di}%
  \BibitemOpen
  \bibfield  {author} {\bibinfo {author} {\bibfnamefont {C.}~\bibnamefont
  {Miquel}}, \bibinfo {author} {\bibfnamefont {J.~P.}\ \bibnamefont {Paz}}, \
  and\ \bibinfo {author} {\bibfnamefont {W.~H.}\ \bibnamefont {Zurek}},\
  }\href@noop {} {\bibfield  {journal} {\bibinfo  {journal} {Phys. Rev. Lett.}\
  }\textbf {\bibinfo {volume} {78}},\ \bibinfo {pages} {3971} (\bibinfo {year}
  {1997})}\BibitemShut {NoStop}%
\bibitem [{\citenamefont {Judson}\ and\ \citenamefont
  {Rabitz}(1992)}]{Judson:1992dg}%
  \BibitemOpen
  \bibfield  {author} {\bibinfo {author} {\bibfnamefont {R.~S.}\ \bibnamefont
  {Judson}}\ and\ \bibinfo {author} {\bibfnamefont {H.}~\bibnamefont
  {Rabitz}},\ }\href@noop {} {\bibfield  {journal} {\bibinfo  {journal} {Phys.
  Rev. Lett.}\ }\textbf {\bibinfo {volume} {68}},\ \bibinfo {pages} {1500}
  (\bibinfo {year} {1992})}\BibitemShut {NoStop}%
\bibitem [{\citenamefont {Rabitz}\ \emph {et~al.}(2000)\citenamefont {Rabitz},
  \citenamefont {de~Vivie-Riedle}, \citenamefont {Motzkus},\ and\ \citenamefont
  {Kompa}}]{Rabitz:2000dn}%
  \BibitemOpen
  \bibfield  {author} {\bibinfo {author} {\bibfnamefont {H.}~\bibnamefont
  {Rabitz}}, \bibinfo {author} {\bibfnamefont {R.}~\bibnamefont
  {de~Vivie-Riedle}}, \bibinfo {author} {\bibfnamefont {M.}~\bibnamefont
  {Motzkus}}, \ and\ \bibinfo {author} {\bibfnamefont {K.}~\bibnamefont
  {Kompa}},\ }\href@noop {} {\bibfield  {journal} {\bibinfo  {journal}
  {Science}\ }\textbf {\bibinfo {volume} {288}},\ \bibinfo {pages} {824}
  (\bibinfo {year} {2000})}\BibitemShut {NoStop}%
\bibitem [{\citenamefont {Ferrie}\ and\ \citenamefont
  {Moussa}(2015)}]{Ferrie:2015fe}%
  \BibitemOpen
  \bibfield  {author} {\bibinfo {author} {\bibfnamefont {C.}~\bibnamefont
  {Ferrie}}\ and\ \bibinfo {author} {\bibfnamefont {O.}~\bibnamefont
  {Moussa}},\ }\href@noop {} {\bibfield  {journal} {\bibinfo  {journal} {Phys.
  Rev. A}\ }\textbf {\bibinfo {volume} {91}},\ \bibinfo {pages} {052306}
  (\bibinfo {year} {2015})}\BibitemShut {NoStop}%
\bibitem [{\citenamefont {James}\ \emph {et~al.}(2001)\citenamefont {James},
  \citenamefont {Kwiat}, \citenamefont {Munro},\ and\ \citenamefont
  {White}}]{James:2001bb}%
  \BibitemOpen
  \bibfield  {author} {\bibinfo {author} {\bibfnamefont {D.~F.~V.}\
  \bibnamefont {James}}, \bibinfo {author} {\bibfnamefont {P.~G.}\ \bibnamefont
  {Kwiat}}, \bibinfo {author} {\bibfnamefont {W.~J.}\ \bibnamefont {Munro}}, \
  and\ \bibinfo {author} {\bibfnamefont {A.~G.}\ \bibnamefont {White}},\
  }\href@noop {} {\ \textbf {\bibinfo {volume} {64}} (\bibinfo {year}
  {2001})}\BibitemShut {NoStop}%
\bibitem [{\citenamefont {O'Brien}\ \emph {et~al.}(2004)\citenamefont
  {O'Brien}, \citenamefont {Pryde}, \citenamefont {Gilchrist}, \citenamefont
  {James}, \citenamefont {Langford}, \citenamefont {Ralph},\ and\ \citenamefont
  {White}}]{OBrien:2004cn}%
  \BibitemOpen
  \bibfield  {author} {\bibinfo {author} {\bibfnamefont {J.~L.}\ \bibnamefont
  {O'Brien}}, \bibinfo {author} {\bibfnamefont {G.}~\bibnamefont {Pryde}},
  \bibinfo {author} {\bibfnamefont {A.}~\bibnamefont {Gilchrist}}, \bibinfo
  {author} {\bibfnamefont {D.~F.~V.}\ \bibnamefont {James}}, \bibinfo {author}
  {\bibfnamefont {N.~K.}\ \bibnamefont {Langford}}, \bibinfo {author}
  {\bibfnamefont {T.~C.}\ \bibnamefont {Ralph}}, \ and\ \bibinfo {author}
  {\bibfnamefont {A.~G.}\ \bibnamefont {White}},\ }\href@noop {} {\bibfield
  {journal} {\bibinfo  {journal} {Phys. Rev. Lett.}\ }\textbf {\bibinfo
  {volume} {93}},\ \bibinfo {pages} {080502} (\bibinfo {year}
  {2004})}\BibitemShut {NoStop}%
\bibitem [{\citenamefont {Egger}\ and\ \citenamefont
  {Wilhelm}(2014)}]{Egger:2014di}%
  \BibitemOpen
  \bibfield  {author} {\bibinfo {author} {\bibfnamefont {D.~J.}\ \bibnamefont
  {Egger}}\ and\ \bibinfo {author} {\bibfnamefont {F.~K.}\ \bibnamefont
  {Wilhelm}},\ }\href@noop {} {\bibfield  {journal} {\bibinfo  {journal} {Phys.
  Rev. Lett.}\ }\textbf {\bibinfo {volume} {112}},\ \bibinfo {pages} {240504}
  (\bibinfo {year} {2014})}\BibitemShut {NoStop}%
\bibitem [{\citenamefont {Kelly}\ \emph {et~al.}(2014)\citenamefont {Kelly},
  \citenamefont {Barends}, \citenamefont {Campbell}, \citenamefont {Chen},
  \citenamefont {Chen}, \citenamefont {Chiaro}, \citenamefont {Dunsworth},
  \citenamefont {Fowler}, \citenamefont {Hoi}, \citenamefont {Jeffrey},
  \citenamefont {Megrant}, \citenamefont {Mutus}, \citenamefont {Neill},
  \citenamefont {O{\textquoteright}Malley}, \citenamefont {Quintana},
  \citenamefont {Roushan}, \citenamefont {Sank}, \citenamefont {Vainsencher},
  \citenamefont {Wenner}, \citenamefont {White}, \citenamefont {Cleland},\ and\
  \citenamefont {Martinis}}]{Kelly:2014fh}%
  \BibitemOpen
  \bibfield  {author} {\bibinfo {author} {\bibfnamefont {J.}~\bibnamefont
  {Kelly}}, \bibinfo {author} {\bibfnamefont {R.}~\bibnamefont {Barends}},
  \bibinfo {author} {\bibfnamefont {B.}~\bibnamefont {Campbell}}, \bibinfo
  {author} {\bibfnamefont {Y.}~\bibnamefont {Chen}}, \bibinfo {author}
  {\bibfnamefont {Z.}~\bibnamefont {Chen}}, \bibinfo {author} {\bibfnamefont
  {B.}~\bibnamefont {Chiaro}}, \bibinfo {author} {\bibfnamefont
  {A.}~\bibnamefont {Dunsworth}}, \bibinfo {author} {\bibfnamefont {A.~G.}\
  \bibnamefont {Fowler}}, \bibinfo {author} {\bibfnamefont {I.~C.}\
  \bibnamefont {Hoi}}, \bibinfo {author} {\bibfnamefont {E.}~\bibnamefont
  {Jeffrey}}, \bibinfo {author} {\bibfnamefont {A.}~\bibnamefont {Megrant}},
  \bibinfo {author} {\bibfnamefont {J.}~\bibnamefont {Mutus}}, \bibinfo
  {author} {\bibfnamefont {C.}~\bibnamefont {Neill}}, \bibinfo {author}
  {\bibfnamefont {P.~J.~J.}\ \bibnamefont {O{\textquoteright}Malley}}, \bibinfo
  {author} {\bibfnamefont {C.}~\bibnamefont {Quintana}}, \bibinfo {author}
  {\bibfnamefont {P.}~\bibnamefont {Roushan}}, \bibinfo {author} {\bibfnamefont
  {D.}~\bibnamefont {Sank}}, \bibinfo {author} {\bibfnamefont {A.}~\bibnamefont
  {Vainsencher}}, \bibinfo {author} {\bibfnamefont {J.}~\bibnamefont {Wenner}},
  \bibinfo {author} {\bibfnamefont {T.~C.}\ \bibnamefont {White}}, \bibinfo
  {author} {\bibfnamefont {A.~N.}\ \bibnamefont {Cleland}}, \ and\ \bibinfo
  {author} {\bibfnamefont {J.~M.}\ \bibnamefont {Martinis}},\ }\href@noop {}
  {\bibfield  {journal} {\bibinfo  {journal} {Phys. Rev. Lett.}\ }\textbf
  {\bibinfo {volume} {112}},\ \bibinfo {pages} {240503} (\bibinfo {year}
  {2014})}\BibitemShut {NoStop}%
\bibitem [{\citenamefont {Kelly}\ \emph {et~al.}(2016)\citenamefont {Kelly},
  \citenamefont {Barends}, \citenamefont {Fowler}, \citenamefont {Megrant},
  \citenamefont {Jeffrey}, \citenamefont {White}, \citenamefont {Sank},
  \citenamefont {Mutus}, \citenamefont {Campbell}, \citenamefont {Chen},
  \citenamefont {Chen}, \citenamefont {Chiaro}, \citenamefont {Dunsworth},
  \citenamefont {Lucero}, \citenamefont {Neeley}, \citenamefont {Neill},
  \citenamefont {O{\textquoteright}Malley}, \citenamefont {Quintana},
  \citenamefont {Roushan}, \citenamefont {Vainsencher}, \citenamefont
  {Wenner},\ and\ \citenamefont {Martinis}}]{Kelly:2016fn}%
  \BibitemOpen
  \bibfield  {author} {\bibinfo {author} {\bibfnamefont {J.}~\bibnamefont
  {Kelly}}, \bibinfo {author} {\bibfnamefont {R.}~\bibnamefont {Barends}},
  \bibinfo {author} {\bibfnamefont {A.~G.}\ \bibnamefont {Fowler}}, \bibinfo
  {author} {\bibfnamefont {A.}~\bibnamefont {Megrant}}, \bibinfo {author}
  {\bibfnamefont {E.}~\bibnamefont {Jeffrey}}, \bibinfo {author} {\bibfnamefont
  {T.~C.}\ \bibnamefont {White}}, \bibinfo {author} {\bibfnamefont
  {D.}~\bibnamefont {Sank}}, \bibinfo {author} {\bibfnamefont {J.~Y.}\
  \bibnamefont {Mutus}}, \bibinfo {author} {\bibfnamefont {B.}~\bibnamefont
  {Campbell}}, \bibinfo {author} {\bibfnamefont {Y.}~\bibnamefont {Chen}},
  \bibinfo {author} {\bibfnamefont {Z.}~\bibnamefont {Chen}}, \bibinfo {author}
  {\bibfnamefont {B.}~\bibnamefont {Chiaro}}, \bibinfo {author} {\bibfnamefont
  {A.}~\bibnamefont {Dunsworth}}, \bibinfo {author} {\bibfnamefont
  {E.}~\bibnamefont {Lucero}}, \bibinfo {author} {\bibfnamefont
  {M.}~\bibnamefont {Neeley}}, \bibinfo {author} {\bibfnamefont
  {C.}~\bibnamefont {Neill}}, \bibinfo {author} {\bibfnamefont {P.~J.~J.}\
  \bibnamefont {O{\textquoteright}Malley}}, \bibinfo {author} {\bibfnamefont
  {C.}~\bibnamefont {Quintana}}, \bibinfo {author} {\bibfnamefont
  {P.}~\bibnamefont {Roushan}}, \bibinfo {author} {\bibfnamefont
  {A.}~\bibnamefont {Vainsencher}}, \bibinfo {author} {\bibfnamefont
  {J.}~\bibnamefont {Wenner}}, \ and\ \bibinfo {author} {\bibfnamefont {J.~M.}\
  \bibnamefont {Martinis}},\ }\href@noop {} {\bibfield  {journal} {\bibinfo
  {journal} {Phys. Rev. A}\ }\textbf {\bibinfo {volume} {94}},\ \bibinfo
  {pages} {032321} (\bibinfo {year} {2016})}\BibitemShut {NoStop}%
\bibitem [{\citenamefont {Lu}\ \emph {et~al.}(2017)\citenamefont {Lu},
  \citenamefont {Li}, \citenamefont {Li}, \citenamefont {Katiyar},
  \citenamefont {Park}, \citenamefont {Feng}, \citenamefont {Xin},
  \citenamefont {Li}, \citenamefont {Long}, \citenamefont {Brodutch},
  \citenamefont {Baugh}, \citenamefont {Zeng},\ and\ \citenamefont
  {Laflamme}}]{Lu:2017ds}%
  \BibitemOpen
  \bibfield  {author} {\bibinfo {author} {\bibfnamefont {D.}~\bibnamefont
  {Lu}}, \bibinfo {author} {\bibfnamefont {K.}~\bibnamefont {Li}}, \bibinfo
  {author} {\bibfnamefont {J.}~\bibnamefont {Li}}, \bibinfo {author}
  {\bibfnamefont {H.}~\bibnamefont {Katiyar}}, \bibinfo {author} {\bibfnamefont
  {A.~J.}\ \bibnamefont {Park}}, \bibinfo {author} {\bibfnamefont
  {G.}~\bibnamefont {Feng}}, \bibinfo {author} {\bibfnamefont {T.}~\bibnamefont
  {Xin}}, \bibinfo {author} {\bibfnamefont {H.}~\bibnamefont {Li}}, \bibinfo
  {author} {\bibfnamefont {G.}~\bibnamefont {Long}}, \bibinfo {author}
  {\bibfnamefont {A.}~\bibnamefont {Brodutch}}, \bibinfo {author}
  {\bibfnamefont {J.}~\bibnamefont {Baugh}}, \bibinfo {author} {\bibfnamefont
  {B.}~\bibnamefont {Zeng}}, \ and\ \bibinfo {author} {\bibfnamefont
  {R.}~\bibnamefont {Laflamme}},\ }\href@noop {} {\bibfield  {journal}
  {\bibinfo  {journal} {npj Quantum Information}\ ,\ \bibinfo {pages} {1}}
  (\bibinfo {year} {2017})}\BibitemShut {NoStop}%
\bibitem [{\citenamefont {Li}\ \emph {et~al.}(2017)\citenamefont {Li},
  \citenamefont {Yang}, \citenamefont {Peng},\ and\ \citenamefont
  {Sun}}]{Li:2017bc}%
  \BibitemOpen
  \bibfield  {author} {\bibinfo {author} {\bibfnamefont {J.}~\bibnamefont
  {Li}}, \bibinfo {author} {\bibfnamefont {X.}~\bibnamefont {Yang}}, \bibinfo
  {author} {\bibfnamefont {X.}~\bibnamefont {Peng}}, \ and\ \bibinfo {author}
  {\bibfnamefont {C.-P.}\ \bibnamefont {Sun}},\ }\href@noop {} {\bibfield
  {journal} {\bibinfo  {journal} {Phys. Rev. Lett.}\ }\textbf {\bibinfo
  {volume} {118}},\ \bibinfo {pages} {150503} (\bibinfo {year}
  {2017})}\BibitemShut {NoStop}%
\bibitem [{\citenamefont {Dive}\ \emph {et~al.}(2017)\citenamefont {Dive},
  \citenamefont {Pitchford}, \citenamefont {Mintert},\ and\ \citenamefont
  {Burgarth}}]{Dive:2017vq}%
  \BibitemOpen
  \bibfield  {author} {\bibinfo {author} {\bibfnamefont {B.}~\bibnamefont
  {Dive}}, \bibinfo {author} {\bibfnamefont {A.}~\bibnamefont {Pitchford}},
  \bibinfo {author} {\bibfnamefont {F.}~\bibnamefont {Mintert}}, \ and\
  \bibinfo {author} {\bibfnamefont {D.}~\bibnamefont {Burgarth}},\ }\href@noop
  {} {\  (\bibinfo {year} {2017})}\BibitemShut {NoStop}%
\bibitem [{\citenamefont {O'Brien}(2007)}]{OBrien:2007ioa}%
  \BibitemOpen
  \bibfield  {author} {\bibinfo {author} {\bibfnamefont {J.~L.}\ \bibnamefont
  {O'Brien}},\ }\href@noop {} {\bibfield  {journal} {\bibinfo  {journal}
  {Science}\ }\textbf {\bibinfo {volume} {318}},\ \bibinfo {pages} {1567}
  (\bibinfo {year} {2007})}\BibitemShut {NoStop}%
\bibitem [{\citenamefont {O'Brien}\ \emph {et~al.}(2009)\citenamefont
  {O'Brien}, \citenamefont {Furusawa},\ and\ \citenamefont {Vu{\v
  c}kovi{\'c}}}]{Obrien:2009eu}%
  \BibitemOpen
  \bibfield  {author} {\bibinfo {author} {\bibfnamefont {J.~L.}\ \bibnamefont
  {O'Brien}}, \bibinfo {author} {\bibfnamefont {A.}~\bibnamefont {Furusawa}}, \
  and\ \bibinfo {author} {\bibfnamefont {J.}~\bibnamefont {Vu{\v
  c}kovi{\'c}}},\ }\href@noop {} {\ \textbf {\bibinfo {volume} {3}},\ \bibinfo
  {pages} {687} (\bibinfo {year} {2009})}\BibitemShut {NoStop}%
\bibitem [{\citenamefont {Silverstone}\ \emph {et~al.}(2014)\citenamefont
  {Silverstone}, \citenamefont {Bonneau}, \citenamefont {Ohira}, \citenamefont
  {Suzuki}, \citenamefont {Yoshida}, \citenamefont {Iizuka}, \citenamefont
  {Ezaki}, \citenamefont {Natarajan}, \citenamefont {Tanner}, \citenamefont
  {Hadfield}, \citenamefont {Zwiller}, \citenamefont {Marshall}, \citenamefont
  {Rarity}, \citenamefont {O'Brien},\ and\ \citenamefont
  {Thompson}}]{Silverstone:2013fu}%
  \BibitemOpen
  \bibfield  {author} {\bibinfo {author} {\bibfnamefont {J.~W.}\ \bibnamefont
  {Silverstone}}, \bibinfo {author} {\bibfnamefont {D.}~\bibnamefont
  {Bonneau}}, \bibinfo {author} {\bibfnamefont {K.}~\bibnamefont {Ohira}},
  \bibinfo {author} {\bibfnamefont {N.}~\bibnamefont {Suzuki}}, \bibinfo
  {author} {\bibfnamefont {H.}~\bibnamefont {Yoshida}}, \bibinfo {author}
  {\bibfnamefont {N.}~\bibnamefont {Iizuka}}, \bibinfo {author} {\bibfnamefont
  {M.}~\bibnamefont {Ezaki}}, \bibinfo {author} {\bibfnamefont {C.~M.}\
  \bibnamefont {Natarajan}}, \bibinfo {author} {\bibfnamefont {M.~G.}\
  \bibnamefont {Tanner}}, \bibinfo {author} {\bibfnamefont {R.~H.}\
  \bibnamefont {Hadfield}}, \bibinfo {author} {\bibfnamefont {V.}~\bibnamefont
  {Zwiller}}, \bibinfo {author} {\bibfnamefont {G.~D.}\ \bibnamefont
  {Marshall}}, \bibinfo {author} {\bibfnamefont {J.~G.}\ \bibnamefont
  {Rarity}}, \bibinfo {author} {\bibfnamefont {J.~L.}\ \bibnamefont {O'Brien}},
  \ and\ \bibinfo {author} {\bibfnamefont {M.~G.}\ \bibnamefont {Thompson}},\
  }\href@noop {} {\bibfield  {journal} {\bibinfo  {journal} {Nat. Photon.}\
  }\textbf {\bibinfo {volume} {8}},\ \bibinfo {pages} {104} (\bibinfo {year}
  {2014})}\BibitemShut {NoStop}%
\bibitem [{\citenamefont {Harris}\ \emph
  {et~al.}(2014{\natexlab{a}})\citenamefont {Harris}, \citenamefont {Grassani},
  \citenamefont {Simbula}, \citenamefont {Pant}, \citenamefont {Galli},
  \citenamefont {Baehr-Jones}, \citenamefont {Hochberg}, \citenamefont
  {Englund}, \citenamefont {Bajoni},\ and\ \citenamefont
  {Galland}}]{Harris:2014kj}%
  \BibitemOpen
  \bibfield  {author} {\bibinfo {author} {\bibfnamefont {N.~C.}\ \bibnamefont
  {Harris}}, \bibinfo {author} {\bibfnamefont {D.}~\bibnamefont {Grassani}},
  \bibinfo {author} {\bibfnamefont {A.}~\bibnamefont {Simbula}}, \bibinfo
  {author} {\bibfnamefont {M.}~\bibnamefont {Pant}}, \bibinfo {author}
  {\bibfnamefont {M.}~\bibnamefont {Galli}}, \bibinfo {author} {\bibfnamefont
  {T.}~\bibnamefont {Baehr-Jones}}, \bibinfo {author} {\bibfnamefont
  {M.}~\bibnamefont {Hochberg}}, \bibinfo {author} {\bibfnamefont
  {D.}~\bibnamefont {Englund}}, \bibinfo {author} {\bibfnamefont
  {D.}~\bibnamefont {Bajoni}}, \ and\ \bibinfo {author} {\bibfnamefont
  {C.}~\bibnamefont {Galland}},\ }\href@noop {} {\bibfield  {journal} {\bibinfo
   {journal} {Phys. Rev. X}\ }\textbf {\bibinfo {volume} {4}},\ \bibinfo
  {pages} {041047} (\bibinfo {year} {2014}{\natexlab{a}})}\BibitemShut
  {NoStop}%
\bibitem [{\citenamefont {Laing}\ \emph {et~al.}(2010)\citenamefont {Laing},
  \citenamefont {Peruzzo}, \citenamefont {Politi}, \citenamefont {Verde},
  \citenamefont {Hadler}, \citenamefont {Ralph}, \citenamefont {Thompson},\
  and\ \citenamefont {O'Brien}}]{Laing:2010fk}%
  \BibitemOpen
  \bibfield  {author} {\bibinfo {author} {\bibfnamefont {A.}~\bibnamefont
  {Laing}}, \bibinfo {author} {\bibfnamefont {A.}~\bibnamefont {Peruzzo}},
  \bibinfo {author} {\bibfnamefont {A.}~\bibnamefont {Politi}}, \bibinfo
  {author} {\bibfnamefont {M.~R.}\ \bibnamefont {Verde}}, \bibinfo {author}
  {\bibfnamefont {M.}~\bibnamefont {Hadler}}, \bibinfo {author} {\bibfnamefont
  {T.~C.}\ \bibnamefont {Ralph}}, \bibinfo {author} {\bibfnamefont {M.~G.}\
  \bibnamefont {Thompson}}, \ and\ \bibinfo {author} {\bibfnamefont {J.~L.}\
  \bibnamefont {O'Brien}},\ }\href@noop {} {\bibfield  {journal} {\bibinfo
  {journal} {App. Phys. Lett.}\ }\textbf {\bibinfo {volume} {97}},\ \bibinfo
  {pages} {211109} (\bibinfo {year} {2010})}\BibitemShut {NoStop}%
\bibitem [{\citenamefont {Carolan}\ \emph {et~al.}(2015)\citenamefont
  {Carolan}, \citenamefont {Harrold}, \citenamefont {Sparrow}, \citenamefont
  {Mart{\'\i}n~L{\'o}pez}, \citenamefont {Russell}, \citenamefont
  {Silverstone}, \citenamefont {Shadbolt}, \citenamefont {Matsuda},
  \citenamefont {Oguma}, \citenamefont {Itoh}, \citenamefont {Marshall},
  \citenamefont {Thompson}, \citenamefont {Matthews}, \citenamefont
  {Hashimoto}, \citenamefont {O'Brien},\ and\ \citenamefont
  {Laing}}]{Carolan:2015vga}%
  \BibitemOpen
  \bibfield  {author} {\bibinfo {author} {\bibfnamefont {J.}~\bibnamefont
  {Carolan}}, \bibinfo {author} {\bibfnamefont {C.}~\bibnamefont {Harrold}},
  \bibinfo {author} {\bibfnamefont {C.}~\bibnamefont {Sparrow}}, \bibinfo
  {author} {\bibfnamefont {E.}~\bibnamefont {Mart{\'\i}n~L{\'o}pez}}, \bibinfo
  {author} {\bibfnamefont {N.~J.}\ \bibnamefont {Russell}}, \bibinfo {author}
  {\bibfnamefont {J.~W.}\ \bibnamefont {Silverstone}}, \bibinfo {author}
  {\bibfnamefont {P.~J.}\ \bibnamefont {Shadbolt}}, \bibinfo {author}
  {\bibfnamefont {N.}~\bibnamefont {Matsuda}}, \bibinfo {author} {\bibfnamefont
  {M.}~\bibnamefont {Oguma}}, \bibinfo {author} {\bibfnamefont
  {M.}~\bibnamefont {Itoh}}, \bibinfo {author} {\bibfnamefont {G.~D.}\
  \bibnamefont {Marshall}}, \bibinfo {author} {\bibfnamefont {M.~G.}\
  \bibnamefont {Thompson}}, \bibinfo {author} {\bibfnamefont {J.~C.~F.}\
  \bibnamefont {Matthews}}, \bibinfo {author} {\bibfnamefont {T.}~\bibnamefont
  {Hashimoto}}, \bibinfo {author} {\bibfnamefont {J.~L.}\ \bibnamefont
  {O'Brien}}, \ and\ \bibinfo {author} {\bibfnamefont {A.}~\bibnamefont
  {Laing}},\ }\href@noop {} {\bibfield  {journal} {\bibinfo  {journal}
  {Science}\ }\textbf {\bibinfo {volume} {349}},\ \bibinfo {pages} {711}
  (\bibinfo {year} {2015})}\BibitemShut {NoStop}%
\bibitem [{\citenamefont {Najafi}\ \emph {et~al.}(2015)\citenamefont {Najafi},
  \citenamefont {Mower}, \citenamefont {Harris}, \citenamefont {Bellei},
  \citenamefont {Dane}, \citenamefont {Lee}, \citenamefont {Hu}, \citenamefont
  {Kharel}, \citenamefont {Marsili}, \citenamefont {Assefa}, \citenamefont
  {Berggren},\ and\ \citenamefont {Englund}}]{Najafi:2015ey}%
  \BibitemOpen
  \bibfield  {author} {\bibinfo {author} {\bibfnamefont {F.}~\bibnamefont
  {Najafi}}, \bibinfo {author} {\bibfnamefont {J.}~\bibnamefont {Mower}},
  \bibinfo {author} {\bibfnamefont {N.~C.}\ \bibnamefont {Harris}}, \bibinfo
  {author} {\bibfnamefont {F.}~\bibnamefont {Bellei}}, \bibinfo {author}
  {\bibfnamefont {A.}~\bibnamefont {Dane}}, \bibinfo {author} {\bibfnamefont
  {C.}~\bibnamefont {Lee}}, \bibinfo {author} {\bibfnamefont {X.}~\bibnamefont
  {Hu}}, \bibinfo {author} {\bibfnamefont {P.}~\bibnamefont {Kharel}}, \bibinfo
  {author} {\bibfnamefont {F.}~\bibnamefont {Marsili}}, \bibinfo {author}
  {\bibfnamefont {S.}~\bibnamefont {Assefa}}, \bibinfo {author} {\bibfnamefont
  {K.~K.}\ \bibnamefont {Berggren}}, \ and\ \bibinfo {author} {\bibfnamefont
  {D.}~\bibnamefont {Englund}},\ }\href@noop {} {\bibfield  {journal} {\bibinfo
   {journal} {Nat. Comms.}\ }\textbf {\bibinfo {volume} {6}},\ \bibinfo {pages}
  {1} (\bibinfo {year} {2015})}\BibitemShut {NoStop}%
\bibitem [{\citenamefont {Silverstone}\ \emph {et~al.}(2016)\citenamefont
  {Silverstone}, \citenamefont {Bonneau}, \citenamefont {O'Brien},\ and\
  \citenamefont {Thompson}}]{Silverstone:2016gha}%
  \BibitemOpen
  \bibfield  {author} {\bibinfo {author} {\bibfnamefont {J.~W.}\ \bibnamefont
  {Silverstone}}, \bibinfo {author} {\bibfnamefont {D.}~\bibnamefont
  {Bonneau}}, \bibinfo {author} {\bibfnamefont {J.~L.}\ \bibnamefont
  {O'Brien}}, \ and\ \bibinfo {author} {\bibfnamefont {M.~G.}\ \bibnamefont
  {Thompson}},\ }\href@noop {} {\bibfield  {journal} {\bibinfo  {journal} {IEEE
  J. Select. Topics Quantum Electron.}\ }\textbf {\bibinfo {volume} {22}},\
  \bibinfo {pages} {1} (\bibinfo {year} {2016})}\BibitemShut {NoStop}%
\bibitem [{\citenamefont {Rudolph}(2017)}]{Rudolph:2017du}%
  \BibitemOpen
  \bibfield  {author} {\bibinfo {author} {\bibfnamefont {T.}~\bibnamefont
  {Rudolph}},\ }\href@noop {} {\bibfield  {journal} {\bibinfo  {journal} {APL
  Photonics}\ }\textbf {\bibinfo {volume} {2}},\ \bibinfo {pages} {030901}
  (\bibinfo {year} {2017})}\BibitemShut {NoStop}%
\bibitem [{\citenamefont {Sun}\ \emph {et~al.}(2015)\citenamefont {Sun},
  \citenamefont {Wade}, \citenamefont {Lee}, \citenamefont {Orcutt},
  \citenamefont {Alloatti}, \citenamefont {Georgas}, \citenamefont {Waterman},
  \citenamefont {Shainline}, \citenamefont {Avizienis}, \citenamefont {Lin},
  \citenamefont {Moss}, \citenamefont {Kumar}, \citenamefont {Pavanello},
  \citenamefont {Atabaki}, \citenamefont {Cook}, \citenamefont {Ou},
  \citenamefont {Leu}, \citenamefont {Chen}, \citenamefont {Asanovi{\'c}},
  \citenamefont {Ram}, \citenamefont {Popovi{\'c}},\ and\ \citenamefont
  {Stojanovi{\'c}}}]{Sun:2015gg}%
  \BibitemOpen
  \bibfield  {author} {\bibinfo {author} {\bibfnamefont {C.}~\bibnamefont
  {Sun}}, \bibinfo {author} {\bibfnamefont {M.~T.}\ \bibnamefont {Wade}},
  \bibinfo {author} {\bibfnamefont {Y.}~\bibnamefont {Lee}}, \bibinfo {author}
  {\bibfnamefont {J.~S.}\ \bibnamefont {Orcutt}}, \bibinfo {author}
  {\bibfnamefont {L.}~\bibnamefont {Alloatti}}, \bibinfo {author}
  {\bibfnamefont {M.~S.}\ \bibnamefont {Georgas}}, \bibinfo {author}
  {\bibfnamefont {A.~S.}\ \bibnamefont {Waterman}}, \bibinfo {author}
  {\bibfnamefont {J.~M.}\ \bibnamefont {Shainline}}, \bibinfo {author}
  {\bibfnamefont {R.~R.}\ \bibnamefont {Avizienis}}, \bibinfo {author}
  {\bibfnamefont {S.}~\bibnamefont {Lin}}, \bibinfo {author} {\bibfnamefont
  {B.~R.}\ \bibnamefont {Moss}}, \bibinfo {author} {\bibfnamefont
  {R.}~\bibnamefont {Kumar}}, \bibinfo {author} {\bibfnamefont
  {F.}~\bibnamefont {Pavanello}}, \bibinfo {author} {\bibfnamefont {A.~H.}\
  \bibnamefont {Atabaki}}, \bibinfo {author} {\bibfnamefont {H.~M.}\
  \bibnamefont {Cook}}, \bibinfo {author} {\bibfnamefont {A.~J.}\ \bibnamefont
  {Ou}}, \bibinfo {author} {\bibfnamefont {J.~C.}\ \bibnamefont {Leu}},
  \bibinfo {author} {\bibfnamefont {Y.-H.}\ \bibnamefont {Chen}}, \bibinfo
  {author} {\bibfnamefont {K.}~\bibnamefont {Asanovi{\'c}}}, \bibinfo {author}
  {\bibfnamefont {R.~J.}\ \bibnamefont {Ram}}, \bibinfo {author} {\bibfnamefont
  {M.~A.}\ \bibnamefont {Popovi{\'c}}}, \ and\ \bibinfo {author} {\bibfnamefont
  {V.~M.}\ \bibnamefont {Stojanovi{\'c}}},\ }\href@noop {} {\bibfield
  {journal} {\bibinfo  {journal} {Nature}\ }\textbf {\bibinfo {volume} {528}},\
  \bibinfo {pages} {534} (\bibinfo {year} {2015})}\BibitemShut {NoStop}%
\bibitem [{\citenamefont {Harris}\ \emph {et~al.}(2017)\citenamefont {Harris},
  \citenamefont {Steinbrecher}, \citenamefont {Prabhu}, \citenamefont {Lahini},
  \citenamefont {Mower}, \citenamefont {Bunandar}, \citenamefont {Chen},
  \citenamefont {Wong}, \citenamefont {Baehr-Jones}, \citenamefont {Hochberg},
  \citenamefont {Lloyd},\ and\ \citenamefont {Englund}}]{Harris:2017hi}%
  \BibitemOpen
  \bibfield  {author} {\bibinfo {author} {\bibfnamefont {N.~C.}\ \bibnamefont
  {Harris}}, \bibinfo {author} {\bibfnamefont {G.~R.}\ \bibnamefont
  {Steinbrecher}}, \bibinfo {author} {\bibfnamefont {M.}~\bibnamefont
  {Prabhu}}, \bibinfo {author} {\bibfnamefont {Y.}~\bibnamefont {Lahini}},
  \bibinfo {author} {\bibfnamefont {J.}~\bibnamefont {Mower}}, \bibinfo
  {author} {\bibfnamefont {D.}~\bibnamefont {Bunandar}}, \bibinfo {author}
  {\bibfnamefont {C.}~\bibnamefont {Chen}}, \bibinfo {author} {\bibfnamefont
  {F.~N.~C.}\ \bibnamefont {Wong}}, \bibinfo {author} {\bibfnamefont
  {T.}~\bibnamefont {Baehr-Jones}}, \bibinfo {author} {\bibfnamefont
  {M.}~\bibnamefont {Hochberg}}, \bibinfo {author} {\bibfnamefont
  {S.}~\bibnamefont {Lloyd}}, \ and\ \bibinfo {author} {\bibfnamefont
  {D.}~\bibnamefont {Englund}},\ }\href@noop {} {\bibfield  {journal} {\bibinfo
   {journal} {Nat. Photon.}\ }\textbf {\bibinfo {volume} {11}},\ \bibinfo
  {pages} {447} (\bibinfo {year} {2017})}\BibitemShut {NoStop}%
\bibitem [{\citenamefont {Wang}\ \emph {et~al.}(2018)\citenamefont {Wang},
  \citenamefont {Paesani}, \citenamefont {Ding}, \citenamefont {Santagati},
  \citenamefont {Skrzypczyk}, \citenamefont {Salavrakos}, \citenamefont {Tura},
  \citenamefont {Augusiak}, \citenamefont {Man{\v c}inska}, \citenamefont
  {Bacco}, \citenamefont {Bonneau}, \citenamefont {Silverstone}, \citenamefont
  {Gong}, \citenamefont {Ac{\'\i}n}, \citenamefont {Rottwitt}, \citenamefont
  {Oxenl{\o}we}, \citenamefont {O'Brien}, \citenamefont {Laing},\ and\
  \citenamefont {Thompson}}]{Wang:2018gh}%
  \BibitemOpen
  \bibfield  {author} {\bibinfo {author} {\bibfnamefont {J.}~\bibnamefont
  {Wang}}, \bibinfo {author} {\bibfnamefont {S.}~\bibnamefont {Paesani}},
  \bibinfo {author} {\bibfnamefont {Y.}~\bibnamefont {Ding}}, \bibinfo {author}
  {\bibfnamefont {R.}~\bibnamefont {Santagati}}, \bibinfo {author}
  {\bibfnamefont {P.}~\bibnamefont {Skrzypczyk}}, \bibinfo {author}
  {\bibfnamefont {A.}~\bibnamefont {Salavrakos}}, \bibinfo {author}
  {\bibfnamefont {J.}~\bibnamefont {Tura}}, \bibinfo {author} {\bibfnamefont
  {R.}~\bibnamefont {Augusiak}}, \bibinfo {author} {\bibfnamefont
  {L.}~\bibnamefont {Man{\v c}inska}}, \bibinfo {author} {\bibfnamefont
  {D.}~\bibnamefont {Bacco}}, \bibinfo {author} {\bibfnamefont
  {D.}~\bibnamefont {Bonneau}}, \bibinfo {author} {\bibfnamefont {J.~W.}\
  \bibnamefont {Silverstone}}, \bibinfo {author} {\bibfnamefont
  {Q.}~\bibnamefont {Gong}}, \bibinfo {author} {\bibfnamefont {A.}~\bibnamefont
  {Ac{\'\i}n}}, \bibinfo {author} {\bibfnamefont {K.}~\bibnamefont {Rottwitt}},
  \bibinfo {author} {\bibfnamefont {L.~K.}\ \bibnamefont {Oxenl{\o}we}},
  \bibinfo {author} {\bibfnamefont {J.~L.}\ \bibnamefont {O'Brien}}, \bibinfo
  {author} {\bibfnamefont {A.}~\bibnamefont {Laing}}, \ and\ \bibinfo {author}
  {\bibfnamefont {M.~G.}\ \bibnamefont {Thompson}},\ }\href@noop {} {\bibfield
  {journal} {\bibinfo  {journal} {Science}\ }\textbf {\bibinfo {volume}
  {360}},\ \bibinfo {pages} {eaar7053} (\bibinfo {year} {2018})}\BibitemShut
  {NoStop}%
\bibitem [{\citenamefont {Rahimi-Keshari}\ \emph {et~al.}(2013)\citenamefont
  {Rahimi-Keshari}, \citenamefont {Broome}, \citenamefont {Fickler},
  \citenamefont {Fedrizzi}, \citenamefont {Ralph},\ and\ \citenamefont
  {White}}]{RahimiKeshari:2013bq}%
  \BibitemOpen
  \bibfield  {author} {\bibinfo {author} {\bibfnamefont {S.}~\bibnamefont
  {Rahimi-Keshari}}, \bibinfo {author} {\bibfnamefont {M.~A.}\ \bibnamefont
  {Broome}}, \bibinfo {author} {\bibfnamefont {R.}~\bibnamefont {Fickler}},
  \bibinfo {author} {\bibfnamefont {A.}~\bibnamefont {Fedrizzi}}, \bibinfo
  {author} {\bibfnamefont {T.~C.}\ \bibnamefont {Ralph}}, \ and\ \bibinfo
  {author} {\bibfnamefont {A.~G.}\ \bibnamefont {White}},\ }\href@noop {}
  {\bibfield  {journal} {\bibinfo  {journal} {Opt. Exp.}\ }\textbf {\bibinfo
  {volume} {21}},\ \bibinfo {pages} {13450} (\bibinfo {year}
  {2013})}\BibitemShut {NoStop}%
\bibitem [{\citenamefont {Grassani}\ \emph {et~al.}(2016)\citenamefont
  {Grassani}, \citenamefont {Simbula}, \citenamefont {Pirotta}, \citenamefont
  {Galli}, \citenamefont {Menotti}, \citenamefont {Harris}, \citenamefont
  {Baehr-Jones}, \citenamefont {Hochberg}, \citenamefont {Galland},
  \citenamefont {Liscidini},\ and\ \citenamefont {Bajoni}}]{Grassani:2016fu}%
  \BibitemOpen
  \bibfield  {author} {\bibinfo {author} {\bibfnamefont {D.}~\bibnamefont
  {Grassani}}, \bibinfo {author} {\bibfnamefont {A.}~\bibnamefont {Simbula}},
  \bibinfo {author} {\bibfnamefont {S.}~\bibnamefont {Pirotta}}, \bibinfo
  {author} {\bibfnamefont {M.}~\bibnamefont {Galli}}, \bibinfo {author}
  {\bibfnamefont {M.}~\bibnamefont {Menotti}}, \bibinfo {author} {\bibfnamefont
  {N.~C.}\ \bibnamefont {Harris}}, \bibinfo {author} {\bibfnamefont
  {T.}~\bibnamefont {Baehr-Jones}}, \bibinfo {author} {\bibfnamefont
  {M.}~\bibnamefont {Hochberg}}, \bibinfo {author} {\bibfnamefont
  {C.}~\bibnamefont {Galland}}, \bibinfo {author} {\bibfnamefont
  {M.}~\bibnamefont {Liscidini}}, \ and\ \bibinfo {author} {\bibfnamefont
  {D.}~\bibnamefont {Bajoni}},\ }\href@noop {} {\bibfield  {journal} {\bibinfo
  {journal} {Sci. Rep.}\ }\textbf {\bibinfo {volume} {6}} (\bibinfo {year}
  {2016})}\BibitemShut {NoStop}%
\bibitem [{\citenamefont {Bogaerts}\ \emph {et~al.}(2011)\citenamefont
  {Bogaerts}, \citenamefont {De~Heyn}, \citenamefont {Van~Vaerenbergh},
  \citenamefont {De~Vos}, \citenamefont {Kumar~Selvaraja}, \citenamefont
  {Claes}, \citenamefont {Dumon}, \citenamefont {Bienstman}, \citenamefont
  {Van~Thourhout},\ and\ \citenamefont {Baets}}]{Bogaerts:2011eha}%
  \BibitemOpen
  \bibfield  {author} {\bibinfo {author} {\bibfnamefont {W.}~\bibnamefont
  {Bogaerts}}, \bibinfo {author} {\bibfnamefont {P.}~\bibnamefont {De~Heyn}},
  \bibinfo {author} {\bibfnamefont {T.}~\bibnamefont {Van~Vaerenbergh}},
  \bibinfo {author} {\bibfnamefont {K.}~\bibnamefont {De~Vos}}, \bibinfo
  {author} {\bibfnamefont {S.}~\bibnamefont {Kumar~Selvaraja}}, \bibinfo
  {author} {\bibfnamefont {T.}~\bibnamefont {Claes}}, \bibinfo {author}
  {\bibfnamefont {P.}~\bibnamefont {Dumon}}, \bibinfo {author} {\bibfnamefont
  {P.}~\bibnamefont {Bienstman}}, \bibinfo {author} {\bibfnamefont
  {D.}~\bibnamefont {Van~Thourhout}}, \ and\ \bibinfo {author} {\bibfnamefont
  {R.}~\bibnamefont {Baets}},\ }\href@noop {} {\bibfield  {journal} {\bibinfo
  {journal} {Laser {\&} Photonics Reviews}\ }\textbf {\bibinfo {volume} {6}},\
  \bibinfo {pages} {47} (\bibinfo {year} {2011})}\BibitemShut {NoStop}%
\bibitem [{\citenamefont {Vernon}\ \emph {et~al.}(2017)\citenamefont {Vernon},
  \citenamefont {Menotti}, \citenamefont {Tison}, \citenamefont {Steidle},
  \citenamefont {Fanto}, \citenamefont {Thomas}, \citenamefont {Preble},
  \citenamefont {Smith}, \citenamefont {Alsing}, \citenamefont {Liscidini},\
  and\ \citenamefont {Sipe}}]{Vernon:2017ei}%
  \BibitemOpen
  \bibfield  {author} {\bibinfo {author} {\bibfnamefont {Z.}~\bibnamefont
  {Vernon}}, \bibinfo {author} {\bibfnamefont {M.}~\bibnamefont {Menotti}},
  \bibinfo {author} {\bibfnamefont {C.~C.}\ \bibnamefont {Tison}}, \bibinfo
  {author} {\bibfnamefont {J.~A.}\ \bibnamefont {Steidle}}, \bibinfo {author}
  {\bibfnamefont {M.~L.}\ \bibnamefont {Fanto}}, \bibinfo {author}
  {\bibfnamefont {P.~M.}\ \bibnamefont {Thomas}}, \bibinfo {author}
  {\bibfnamefont {S.~F.}\ \bibnamefont {Preble}}, \bibinfo {author}
  {\bibfnamefont {A.~M.}\ \bibnamefont {Smith}}, \bibinfo {author}
  {\bibfnamefont {P.~M.}\ \bibnamefont {Alsing}}, \bibinfo {author}
  {\bibfnamefont {M.}~\bibnamefont {Liscidini}}, \ and\ \bibinfo {author}
  {\bibfnamefont {J.~E.}\ \bibnamefont {Sipe}},\ }\href@noop {} {\bibfield
  {journal} {\bibinfo  {journal} {Opt. Lett., OL}\ }\textbf {\bibinfo {volume}
  {42}},\ \bibinfo {pages} {3638} (\bibinfo {year} {2017})}\BibitemShut
  {NoStop}%
\bibitem [{\citenamefont {Aaronson}\ and\ \citenamefont
  {Arkhipov}(2011)}]{Aaronson:2011tja}%
  \BibitemOpen
  \bibfield  {author} {\bibinfo {author} {\bibfnamefont {S.}~\bibnamefont
  {Aaronson}}\ and\ \bibinfo {author} {\bibfnamefont {A.}~\bibnamefont
  {Arkhipov}},\ }in\ \href@noop {} {\emph {\bibinfo {booktitle} {Proceedings of
  the 43rd annual ACM symposium on Theory of computing}}}\ (\bibinfo
  {publisher} {ACM},\ \bibinfo {year} {2011})\ pp.\ \bibinfo {pages}
  {333--342}\BibitemShut {NoStop}%
\bibitem [{\citenamefont {Neville}\ \emph {et~al.}(2017)\citenamefont
  {Neville}, \citenamefont {Sparrow}, \citenamefont {Clifford}, \citenamefont
  {Johnston}, \citenamefont {Birchall}, \citenamefont {Montanaro},\ and\
  \citenamefont {Laing}}]{Neville:2017do}%
  \BibitemOpen
  \bibfield  {author} {\bibinfo {author} {\bibfnamefont {A.}~\bibnamefont
  {Neville}}, \bibinfo {author} {\bibfnamefont {C.}~\bibnamefont {Sparrow}},
  \bibinfo {author} {\bibfnamefont {R.}~\bibnamefont {Clifford}}, \bibinfo
  {author} {\bibfnamefont {E.}~\bibnamefont {Johnston}}, \bibinfo {author}
  {\bibfnamefont {P.~M.}\ \bibnamefont {Birchall}}, \bibinfo {author}
  {\bibfnamefont {A.}~\bibnamefont {Montanaro}}, \ and\ \bibinfo {author}
  {\bibfnamefont {A.}~\bibnamefont {Laing}},\ }\href@noop {} {\bibfield
  {journal} {\bibinfo  {journal} {Nat. Phys.}\ }\textbf {\bibinfo {volume}
  {13}},\ \bibinfo {pages} {1153} (\bibinfo {year} {2017})}\BibitemShut
  {NoStop}%
\bibitem [{\citenamefont {Aspuru-Guzik}\ and\ \citenamefont
  {Walther}(2012)}]{AspuruGuzik:2012ho}%
  \BibitemOpen
  \bibfield  {author} {\bibinfo {author} {\bibfnamefont {A.}~\bibnamefont
  {Aspuru-Guzik}}\ and\ \bibinfo {author} {\bibfnamefont {P.}~\bibnamefont
  {Walther}},\ }\href@noop {} {\bibfield  {journal} {\bibinfo  {journal} {Nat.
  Phys.}\ }\textbf {\bibinfo {volume} {8}},\ \bibinfo {pages} {285} (\bibinfo
  {year} {2012})}\BibitemShut {NoStop}%
\bibitem [{\citenamefont {Huh}\ \emph {et~al.}(2015)\citenamefont {Huh},
  \citenamefont {Guerreschi}, \citenamefont {Peropadre}, \citenamefont
  {McClean},\ and\ \citenamefont {Aspuru-Guzik}}]{Huh:2014vk}%
  \BibitemOpen
  \bibfield  {author} {\bibinfo {author} {\bibfnamefont {J.}~\bibnamefont
  {Huh}}, \bibinfo {author} {\bibfnamefont {G.~G.}\ \bibnamefont {Guerreschi}},
  \bibinfo {author} {\bibfnamefont {B.}~\bibnamefont {Peropadre}}, \bibinfo
  {author} {\bibfnamefont {J.~R.}\ \bibnamefont {McClean}}, \ and\ \bibinfo
  {author} {\bibfnamefont {A.}~\bibnamefont {Aspuru-Guzik}},\ }\href@noop {}
  {\bibfield  {journal} {\bibinfo  {journal} {Nat. Photon.}\ }\textbf {\bibinfo
  {volume} {9}},\ \bibinfo {pages} {615} (\bibinfo {year} {2015})}\BibitemShut
  {NoStop}%
\bibitem [{\citenamefont {Sparrow}\ \emph {et~al.}(2018)\citenamefont
  {Sparrow}, \citenamefont {Mart{\'\i}n~L{\'o}pez}, \citenamefont {Maraviglia},
  \citenamefont {Neville}, \citenamefont {Harrold}, \citenamefont {Carolan},
  \citenamefont {Joglekar}, \citenamefont {Hashimoto}, \citenamefont {Matsuda},
  \citenamefont {O'Brien}, \citenamefont {Tew},\ and\ \citenamefont
  {Laing}}]{Sparrow:2018ba}%
  \BibitemOpen
  \bibfield  {author} {\bibinfo {author} {\bibfnamefont {C.}~\bibnamefont
  {Sparrow}}, \bibinfo {author} {\bibfnamefont {E.}~\bibnamefont
  {Mart{\'\i}n~L{\'o}pez}}, \bibinfo {author} {\bibfnamefont {N.}~\bibnamefont
  {Maraviglia}}, \bibinfo {author} {\bibfnamefont {A.}~\bibnamefont {Neville}},
  \bibinfo {author} {\bibfnamefont {C.}~\bibnamefont {Harrold}}, \bibinfo
  {author} {\bibfnamefont {J.}~\bibnamefont {Carolan}}, \bibinfo {author}
  {\bibfnamefont {Y.~N.}\ \bibnamefont {Joglekar}}, \bibinfo {author}
  {\bibfnamefont {T.}~\bibnamefont {Hashimoto}}, \bibinfo {author}
  {\bibfnamefont {N.}~\bibnamefont {Matsuda}}, \bibinfo {author} {\bibfnamefont
  {J.~L.}\ \bibnamefont {O'Brien}}, \bibinfo {author} {\bibfnamefont {D.~P.}\
  \bibnamefont {Tew}}, \ and\ \bibinfo {author} {\bibfnamefont
  {A.}~\bibnamefont {Laing}},\ }\href@noop {} {\bibfield  {journal} {\bibinfo
  {journal} {Nature}\ }\textbf {\bibinfo {volume} {557}},\ \bibinfo {pages}
  {660} (\bibinfo {year} {2018})}\BibitemShut {NoStop}%
\bibitem [{\citenamefont {Gimeno-Segovia}\ \emph {et~al.}(2015)\citenamefont
  {Gimeno-Segovia}, \citenamefont {Shadbolt}, \citenamefont {Browne},\ and\
  \citenamefont {Rudolph}}]{GimenoSegovia:2015di}%
  \BibitemOpen
  \bibfield  {author} {\bibinfo {author} {\bibfnamefont {M.}~\bibnamefont
  {Gimeno-Segovia}}, \bibinfo {author} {\bibfnamefont {P.}~\bibnamefont
  {Shadbolt}}, \bibinfo {author} {\bibfnamefont {D.~E.}\ \bibnamefont
  {Browne}}, \ and\ \bibinfo {author} {\bibfnamefont {T.}~\bibnamefont
  {Rudolph}},\ }\href@noop {} {\bibfield  {journal} {\bibinfo  {journal} {Phys.
  Rev. Lett.}\ }\textbf {\bibinfo {volume} {115}},\ \bibinfo {pages} {020502}
  (\bibinfo {year} {2015})}\BibitemShut {NoStop}%
\bibitem [{\citenamefont {Rohde}\ and\ \citenamefont
  {Ralph}(2012)}]{Rohde:2012cna}%
  \BibitemOpen
  \bibfield  {author} {\bibinfo {author} {\bibfnamefont {P.~P.}\ \bibnamefont
  {Rohde}}\ and\ \bibinfo {author} {\bibfnamefont {T.~C.}\ \bibnamefont
  {Ralph}},\ }\href@noop {} {\bibfield  {journal} {\bibinfo  {journal} {Phys.
  Rev. A}\ }\textbf {\bibinfo {volume} {85}},\ \bibinfo {pages} {022332}
  (\bibinfo {year} {2012})}\BibitemShut {NoStop}%
\bibitem [{\citenamefont {Shchesnovich}(2014)}]{Shchesnovich:2014ii}%
  \BibitemOpen
  \bibfield  {author} {\bibinfo {author} {\bibfnamefont {V.~S.}\ \bibnamefont
  {Shchesnovich}},\ }\href@noop {} {\bibfield  {journal} {\bibinfo  {journal}
  {Phys. Rev. A}\ }\textbf {\bibinfo {volume} {89}},\ \bibinfo {pages} {022333}
  (\bibinfo {year} {2014})}\BibitemShut {NoStop}%
\bibitem [{\citenamefont {Vernon}\ and\ \citenamefont
  {Sipe}(2015)}]{Vernon:2015fo}%
  \BibitemOpen
  \bibfield  {author} {\bibinfo {author} {\bibfnamefont {Z.}~\bibnamefont
  {Vernon}}\ and\ \bibinfo {author} {\bibfnamefont {J.~E.}\ \bibnamefont
  {Sipe}},\ }\href@noop {} {\bibfield  {journal} {\bibinfo  {journal} {Phys.
  Rev. A}\ }\textbf {\bibinfo {volume} {92}},\ \bibinfo {pages} {033840}
  (\bibinfo {year} {2015})}\BibitemShut {NoStop}%
\bibitem [{\citenamefont {Padmaraju}\ \emph {et~al.}(2013)\citenamefont
  {Padmaraju}, \citenamefont {Logan}, \citenamefont {Shiraishi}, \citenamefont
  {Ackert}, \citenamefont {Knights},\ and\ \citenamefont
  {Bergman}}]{Padmaraju:2013ba}%
  \BibitemOpen
  \bibfield  {author} {\bibinfo {author} {\bibfnamefont {K.}~\bibnamefont
  {Padmaraju}}, \bibinfo {author} {\bibfnamefont {D.~F.}\ \bibnamefont
  {Logan}}, \bibinfo {author} {\bibfnamefont {T.}~\bibnamefont {Shiraishi}},
  \bibinfo {author} {\bibfnamefont {J.~J.}\ \bibnamefont {Ackert}}, \bibinfo
  {author} {\bibfnamefont {A.~P.}\ \bibnamefont {Knights}}, \ and\ \bibinfo
  {author} {\bibfnamefont {K.}~\bibnamefont {Bergman}},\ }\href@noop {}
  {\bibfield  {journal} {\bibinfo  {journal} {Lightwave Technology, Journal
  of}\ }\textbf {\bibinfo {volume} {32}},\ \bibinfo {pages} {505} (\bibinfo
  {year} {2013})}\BibitemShut {NoStop}%
\bibitem [{\citenamefont {Horst}\ \emph {et~al.}(2013)\citenamefont {Horst},
  \citenamefont {Green}, \citenamefont {Assefa}, \citenamefont {Shank},
  \citenamefont {Vlasov},\ and\ \citenamefont {Offrein}}]{Horst:2013il}%
  \BibitemOpen
  \bibfield  {author} {\bibinfo {author} {\bibfnamefont {F.}~\bibnamefont
  {Horst}}, \bibinfo {author} {\bibfnamefont {W.~M.~J.}\ \bibnamefont {Green}},
  \bibinfo {author} {\bibfnamefont {S.}~\bibnamefont {Assefa}}, \bibinfo
  {author} {\bibfnamefont {S.~M.}\ \bibnamefont {Shank}}, \bibinfo {author}
  {\bibfnamefont {Y.~A.}\ \bibnamefont {Vlasov}}, \ and\ \bibinfo {author}
  {\bibfnamefont {B.~J.}\ \bibnamefont {Offrein}},\ }\href@noop {} {\bibfield
  {journal} {\bibinfo  {journal} {Opt. Exp.}\ }\textbf {\bibinfo {volume}
  {21}},\ \bibinfo {pages} {11652} (\bibinfo {year} {2013})}\BibitemShut
  {NoStop}%
\bibitem [{\citenamefont {Michel}\ \emph {et~al.}(2010)\citenamefont {Michel},
  \citenamefont {Liu},\ and\ \citenamefont {Kimerling}}]{Michel:2010ib}%
  \BibitemOpen
  \bibfield  {author} {\bibinfo {author} {\bibfnamefont {J.}~\bibnamefont
  {Michel}}, \bibinfo {author} {\bibfnamefont {J.}~\bibnamefont {Liu}}, \ and\
  \bibinfo {author} {\bibfnamefont {L.~C.}\ \bibnamefont {Kimerling}},\
  }\href@noop {} {\ \textbf {\bibinfo {volume} {4}},\ \bibinfo {pages} {527}
  (\bibinfo {year} {2010})}\BibitemShut {NoStop}%
\bibitem [{\citenamefont {Harris}\ \emph
  {et~al.}(2014{\natexlab{b}})\citenamefont {Harris}, \citenamefont {Ma},
  \citenamefont {Mower}, \citenamefont {Baehr-Jones}, \citenamefont {Englund},
  \citenamefont {Hochberg},\ and\ \citenamefont {Galland}}]{Harris:2014kz}%
  \BibitemOpen
  \bibfield  {author} {\bibinfo {author} {\bibfnamefont {N.~C.}\ \bibnamefont
  {Harris}}, \bibinfo {author} {\bibfnamefont {Y.}~\bibnamefont {Ma}}, \bibinfo
  {author} {\bibfnamefont {J.}~\bibnamefont {Mower}}, \bibinfo {author}
  {\bibfnamefont {T.}~\bibnamefont {Baehr-Jones}}, \bibinfo {author}
  {\bibfnamefont {D.}~\bibnamefont {Englund}}, \bibinfo {author} {\bibfnamefont
  {M.}~\bibnamefont {Hochberg}}, \ and\ \bibinfo {author} {\bibfnamefont
  {C.}~\bibnamefont {Galland}},\ }\href@noop {} {\bibfield  {journal} {\bibinfo
   {journal} {Opt. Exp.}\ }\textbf {\bibinfo {volume} {22}},\ \bibinfo {pages}
  {10487} (\bibinfo {year} {2014}{\natexlab{b}})}\BibitemShut {NoStop}%
\bibitem [{\citenamefont {Seok}\ \emph {et~al.}(2016)\citenamefont {Seok},
  \citenamefont {Quack}, \citenamefont {Han}, \citenamefont {Muller},\ and\
  \citenamefont {Wu}}]{Seok:2016he}%
  \BibitemOpen
  \bibfield  {author} {\bibinfo {author} {\bibfnamefont {T.~J.}\ \bibnamefont
  {Seok}}, \bibinfo {author} {\bibfnamefont {N.}~\bibnamefont {Quack}},
  \bibinfo {author} {\bibfnamefont {S.}~\bibnamefont {Han}}, \bibinfo {author}
  {\bibfnamefont {R.~S.}\ \bibnamefont {Muller}}, \ and\ \bibinfo {author}
  {\bibfnamefont {M.~C.}\ \bibnamefont {Wu}},\ }\href@noop {} {\bibfield
  {journal} {\bibinfo  {journal} {Optica}\ }\textbf {\bibinfo {volume} {3}},\
  \bibinfo {pages} {64} (\bibinfo {year} {2016})}\BibitemShut {NoStop}%
\bibitem [{\citenamefont {Thomson}\ \emph {et~al.}(2011)\citenamefont
  {Thomson}, \citenamefont {Gardes}, \citenamefont {Hu}, \citenamefont
  {Mashanovich}, \citenamefont {Fournier}, \citenamefont {Grosse},
  \citenamefont {Fedeli},\ and\ \citenamefont {Reed}}]{Thomson:2011hk}%
  \BibitemOpen
  \bibfield  {author} {\bibinfo {author} {\bibfnamefont {D.~J.}\ \bibnamefont
  {Thomson}}, \bibinfo {author} {\bibfnamefont {F.~Y.}\ \bibnamefont {Gardes}},
  \bibinfo {author} {\bibfnamefont {Y.}~\bibnamefont {Hu}}, \bibinfo {author}
  {\bibfnamefont {G.}~\bibnamefont {Mashanovich}}, \bibinfo {author}
  {\bibfnamefont {M.}~\bibnamefont {Fournier}}, \bibinfo {author}
  {\bibfnamefont {P.}~\bibnamefont {Grosse}}, \bibinfo {author} {\bibfnamefont
  {J.-M.}\ \bibnamefont {Fedeli}}, \ and\ \bibinfo {author} {\bibfnamefont
  {G.~T.}\ \bibnamefont {Reed}},\ }\href@noop {} {\bibfield  {journal}
  {\bibinfo  {journal} {Opt. Exp.}\ }\textbf {\bibinfo {volume} {19}},\
  \bibinfo {pages} {11507} (\bibinfo {year} {2011})}\BibitemShut {NoStop}%
\bibitem [{\citenamefont {Little}\ \emph {et~al.}(1997)\citenamefont {Little},
  \citenamefont {Laine},\ and\ \citenamefont {Chu}}]{Little:1997kk}%
  \BibitemOpen
  \bibfield  {author} {\bibinfo {author} {\bibfnamefont {B.~E.}\ \bibnamefont
  {Little}}, \bibinfo {author} {\bibfnamefont {J.-P.}\ \bibnamefont {Laine}}, \
  and\ \bibinfo {author} {\bibfnamefont {S.~T.}\ \bibnamefont {Chu}},\
  }\href@noop {} {\bibfield  {journal} {\bibinfo  {journal} {Opt. Lett., OL}\
  }\textbf {\bibinfo {volume} {22}},\ \bibinfo {pages} {4} (\bibinfo {year}
  {1997})}\BibitemShut {NoStop}%
\bibitem [{\citenamefont {Zhou}\ \emph {et~al.}(2015)\citenamefont {Zhou},
  \citenamefont {Yin},\ and\ \citenamefont {Michel}}]{Zhou:2015gj}%
  \BibitemOpen
  \bibfield  {author} {\bibinfo {author} {\bibfnamefont {Z.}~\bibnamefont
  {Zhou}}, \bibinfo {author} {\bibfnamefont {B.}~\bibnamefont {Yin}}, \ and\
  \bibinfo {author} {\bibfnamefont {J.}~\bibnamefont {Michel}},\ }\href@noop {}
  {\bibfield  {journal} {\bibinfo  {journal} {Light Sci Appl}\ }\textbf
  {\bibinfo {volume} {4}},\ \bibinfo {pages} {e358} (\bibinfo {year}
  {2015})}\BibitemShut {NoStop}%
\bibitem [{\citenamefont {Piekarek}\ \emph {et~al.}(2017)\citenamefont
  {Piekarek}, \citenamefont {Bonneau}, \citenamefont {Miki}, \citenamefont
  {Yamashita}, \citenamefont {Fujiwara}, \citenamefont {Sasaki}, \citenamefont
  {Terai}, \citenamefont {Tanner}, \citenamefont {Natarajan}, \citenamefont
  {Hadfield}, \citenamefont {O'Brien},\ and\ \citenamefont
  {Thompson}}]{Piekarek:2017ch}%
  \BibitemOpen
  \bibfield  {author} {\bibinfo {author} {\bibfnamefont {M.}~\bibnamefont
  {Piekarek}}, \bibinfo {author} {\bibfnamefont {D.}~\bibnamefont {Bonneau}},
  \bibinfo {author} {\bibfnamefont {S.}~\bibnamefont {Miki}}, \bibinfo {author}
  {\bibfnamefont {T.}~\bibnamefont {Yamashita}}, \bibinfo {author}
  {\bibfnamefont {M.}~\bibnamefont {Fujiwara}}, \bibinfo {author}
  {\bibfnamefont {M.}~\bibnamefont {Sasaki}}, \bibinfo {author} {\bibfnamefont
  {H.}~\bibnamefont {Terai}}, \bibinfo {author} {\bibfnamefont {M.~G.}\
  \bibnamefont {Tanner}}, \bibinfo {author} {\bibfnamefont {C.~M.}\
  \bibnamefont {Natarajan}}, \bibinfo {author} {\bibfnamefont {R.~H.}\
  \bibnamefont {Hadfield}}, \bibinfo {author} {\bibfnamefont {J.~L.}\
  \bibnamefont {O'Brien}}, \ and\ \bibinfo {author} {\bibfnamefont {M.~G.}\
  \bibnamefont {Thompson}},\ }\href@noop {} {\bibfield  {journal} {\bibinfo
  {journal} {Opt. Lett., OL}\ }\textbf {\bibinfo {volume} {42}},\ \bibinfo
  {pages} {815} (\bibinfo {year} {2017})}\BibitemShut {NoStop}%
\bibitem [{\citenamefont {Wang}\ \emph {et~al.}(2016)\citenamefont {Wang},
  \citenamefont {Liu}, \citenamefont {Sun}, \citenamefont {Huang},
  \citenamefont {Li},\ and\ \citenamefont {Han}}]{Wang:2016jma}%
  \BibitemOpen
  \bibfield  {author} {\bibinfo {author} {\bibfnamefont {Z.}~\bibnamefont
  {Wang}}, \bibinfo {author} {\bibfnamefont {H.}~\bibnamefont {Liu}}, \bibinfo
  {author} {\bibfnamefont {Q.}~\bibnamefont {Sun}}, \bibinfo {author}
  {\bibfnamefont {N.}~\bibnamefont {Huang}}, \bibinfo {author} {\bibfnamefont
  {S.}~\bibnamefont {Li}}, \ and\ \bibinfo {author} {\bibfnamefont
  {J.}~\bibnamefont {Han}},\ }\href@noop {} {\bibfield  {journal} {\bibinfo
  {journal} {Laser Phys.}\ }\textbf {\bibinfo {volume} {26}},\ \bibinfo {pages}
  {075403} (\bibinfo {year} {2016})}\BibitemShut {NoStop}%
\bibitem [{\citenamefont {Lin}\ \emph {et~al.}(2017)\citenamefont {Lin},
  \citenamefont {Luo}, \citenamefont {Gu}, \citenamefont {Kimerling},
  \citenamefont {Wada}, \citenamefont {Agarwal},\ and\ \citenamefont
  {Hu}}]{Lin:2017io}%
  \BibitemOpen
  \bibfield  {author} {\bibinfo {author} {\bibfnamefont {H.}~\bibnamefont
  {Lin}}, \bibinfo {author} {\bibfnamefont {Z.}~\bibnamefont {Luo}}, \bibinfo
  {author} {\bibfnamefont {T.}~\bibnamefont {Gu}}, \bibinfo {author}
  {\bibfnamefont {L.~C.}\ \bibnamefont {Kimerling}}, \bibinfo {author}
  {\bibfnamefont {K.}~\bibnamefont {Wada}}, \bibinfo {author} {\bibfnamefont
  {A.}~\bibnamefont {Agarwal}}, \ and\ \bibinfo {author} {\bibfnamefont
  {J.}~\bibnamefont {Hu}},\ }\href@noop {} {\bibfield  {journal} {\bibinfo
  {journal} {Nanophotonics}\ }\textbf {\bibinfo {volume} {7}},\ \bibinfo
  {pages} {2733} (\bibinfo {year} {2017})}\BibitemShut {NoStop}%
\bibitem [{\citenamefont {Zou}\ \emph {et~al.}(2018)\citenamefont {Zou},
  \citenamefont {Chakravarty}, \citenamefont {Chung}, \citenamefont {Xu},\ and\
  \citenamefont {Chen}}]{Zou:2018ke}%
  \BibitemOpen
  \bibfield  {author} {\bibinfo {author} {\bibfnamefont {Y.}~\bibnamefont
  {Zou}}, \bibinfo {author} {\bibfnamefont {S.}~\bibnamefont {Chakravarty}},
  \bibinfo {author} {\bibfnamefont {C.-J.}\ \bibnamefont {Chung}}, \bibinfo
  {author} {\bibfnamefont {X.}~\bibnamefont {Xu}}, \ and\ \bibinfo {author}
  {\bibfnamefont {R.~T.}\ \bibnamefont {Chen}},\ }\href@noop {} {\bibfield
  {journal} {\bibinfo  {journal} {Photon. Res., PRJ}\ }\textbf {\bibinfo
  {volume} {6}},\ \bibinfo {pages} {254} (\bibinfo {year} {2018})}\BibitemShut
  {NoStop}%
\bibitem [{\citenamefont {Atabaki}\ \emph {et~al.}(2018)\citenamefont
  {Atabaki}, \citenamefont {Moazeni}, \citenamefont {Pavanello}, \citenamefont
  {Gevorgyan}, \citenamefont {Notaros}, \citenamefont {Alloatti}, \citenamefont
  {Wade}, \citenamefont {Sun}, \citenamefont {Kruger}, \citenamefont {Meng},
  \citenamefont {Al~Qubaisi}, \citenamefont {Wang}, \citenamefont {Zhang},
  \citenamefont {Khilo}, \citenamefont {Baiocco}, \citenamefont {Popovi{\'c}},
  \citenamefont {Stojanovi{\'c}},\ and\ \citenamefont {Ram}}]{Atabaki:2018jf}%
  \BibitemOpen
  \bibfield  {author} {\bibinfo {author} {\bibfnamefont {A.~H.}\ \bibnamefont
  {Atabaki}}, \bibinfo {author} {\bibfnamefont {S.}~\bibnamefont {Moazeni}},
  \bibinfo {author} {\bibfnamefont {F.}~\bibnamefont {Pavanello}}, \bibinfo
  {author} {\bibfnamefont {H.}~\bibnamefont {Gevorgyan}}, \bibinfo {author}
  {\bibfnamefont {J.}~\bibnamefont {Notaros}}, \bibinfo {author} {\bibfnamefont
  {L.}~\bibnamefont {Alloatti}}, \bibinfo {author} {\bibfnamefont {M.~T.}\
  \bibnamefont {Wade}}, \bibinfo {author} {\bibfnamefont {C.}~\bibnamefont
  {Sun}}, \bibinfo {author} {\bibfnamefont {S.~A.}\ \bibnamefont {Kruger}},
  \bibinfo {author} {\bibfnamefont {H.}~\bibnamefont {Meng}}, \bibinfo {author}
  {\bibfnamefont {K.}~\bibnamefont {Al~Qubaisi}}, \bibinfo {author}
  {\bibfnamefont {I.}~\bibnamefont {Wang}}, \bibinfo {author} {\bibfnamefont
  {B.}~\bibnamefont {Zhang}}, \bibinfo {author} {\bibfnamefont
  {A.}~\bibnamefont {Khilo}}, \bibinfo {author} {\bibfnamefont {C.~V.}\
  \bibnamefont {Baiocco}}, \bibinfo {author} {\bibfnamefont {M.~A.}\
  \bibnamefont {Popovi{\'c}}}, \bibinfo {author} {\bibfnamefont {V.~M.}\
  \bibnamefont {Stojanovi{\'c}}}, \ and\ \bibinfo {author} {\bibfnamefont
  {R.~J.}\ \bibnamefont {Ram}},\ }\href@noop {} {\bibfield  {journal} {\bibinfo
   {journal} {Nature}\ }\textbf {\bibinfo {volume} {556}},\ \bibinfo {pages}
  {349} (\bibinfo {year} {2018})}\BibitemShut {NoStop}%
\bibitem [{\citenamefont {Carroll}\ \emph {et~al.}(2016)\citenamefont
  {Carroll}, \citenamefont {Lee}, \citenamefont {Scarcella}, \citenamefont
  {Gradkowski}, \citenamefont {Duperron}, \citenamefont {Lu}, \citenamefont
  {Zhao}, \citenamefont {Eason}, \citenamefont {Morrissey}, \citenamefont
  {Rensing}, \citenamefont {Collins}, \citenamefont {Hwang},\ and\
  \citenamefont {O{\textquoteright}Brien}}]{Carroll:2016fa}%
  \BibitemOpen
  \bibfield  {author} {\bibinfo {author} {\bibfnamefont {L.}~\bibnamefont
  {Carroll}}, \bibinfo {author} {\bibfnamefont {J.-S.}\ \bibnamefont {Lee}},
  \bibinfo {author} {\bibfnamefont {C.}~\bibnamefont {Scarcella}}, \bibinfo
  {author} {\bibfnamefont {K.}~\bibnamefont {Gradkowski}}, \bibinfo {author}
  {\bibfnamefont {M.}~\bibnamefont {Duperron}}, \bibinfo {author}
  {\bibfnamefont {H.}~\bibnamefont {Lu}}, \bibinfo {author} {\bibfnamefont
  {Y.}~\bibnamefont {Zhao}}, \bibinfo {author} {\bibfnamefont {C.}~\bibnamefont
  {Eason}}, \bibinfo {author} {\bibfnamefont {P.}~\bibnamefont {Morrissey}},
  \bibinfo {author} {\bibfnamefont {M.}~\bibnamefont {Rensing}}, \bibinfo
  {author} {\bibfnamefont {S.}~\bibnamefont {Collins}}, \bibinfo {author}
  {\bibfnamefont {H.}~\bibnamefont {Hwang}}, \ and\ \bibinfo {author}
  {\bibfnamefont {P.}~\bibnamefont {O{\textquoteright}Brien}},\ }\href@noop {}
  {\bibfield  {journal} {\bibinfo  {journal} {Applied Sciences}\ }\textbf
  {\bibinfo {volume} {6}},\ \bibinfo {pages} {426} (\bibinfo {year}
  {2016})}\BibitemShut {NoStop}%
\bibitem [{\citenamefont {Schuck}\ \emph {et~al.}(2013)\citenamefont {Schuck},
  \citenamefont {Pernice},\ and\ \citenamefont {Tang}}]{Schuck:2013be}%
  \BibitemOpen
  \bibfield  {author} {\bibinfo {author} {\bibfnamefont {C.}~\bibnamefont
  {Schuck}}, \bibinfo {author} {\bibfnamefont {W.~H.~P.}\ \bibnamefont
  {Pernice}}, \ and\ \bibinfo {author} {\bibfnamefont {H.~X.}\ \bibnamefont
  {Tang}},\ }\href@noop {} {\bibfield  {journal} {\bibinfo  {journal} {Sci.
  Rep.}\ }\textbf {\bibinfo {volume} {3}},\ \bibinfo {pages} {696} (\bibinfo
  {year} {2013})}\BibitemShut {NoStop}%
\bibitem [{\citenamefont {Schelew}\ \emph {et~al.}(2015)\citenamefont
  {Schelew}, \citenamefont {Akhlaghi},\ and\ \citenamefont
  {Young}}]{Schelew:2015jw}%
  \BibitemOpen
  \bibfield  {author} {\bibinfo {author} {\bibfnamefont {E.}~\bibnamefont
  {Schelew}}, \bibinfo {author} {\bibfnamefont {M.~K.}\ \bibnamefont
  {Akhlaghi}}, \ and\ \bibinfo {author} {\bibfnamefont {J.~F.}\ \bibnamefont
  {Young}},\ }\href@noop {} {\bibfield  {journal} {\bibinfo  {journal} {Nat.
  Comms.}\ }\textbf {\bibinfo {volume} {6}},\ \bibinfo {pages} {1} (\bibinfo
  {year} {2015})}\BibitemShut {NoStop}%
\bibitem [{\citenamefont {Zhu}\ \emph {et~al.}(2018)\citenamefont {Zhu},
  \citenamefont {Zhao}, \citenamefont {Choi}, \citenamefont {Lu}, \citenamefont
  {Dane}, \citenamefont {Englund},\ and\ \citenamefont
  {Berggren}}]{DiZhu:2018ja}%
  \BibitemOpen
  \bibfield  {author} {\bibinfo {author} {\bibfnamefont {D.}~\bibnamefont
  {Zhu}}, \bibinfo {author} {\bibfnamefont {Q.-Y.}\ \bibnamefont {Zhao}},
  \bibinfo {author} {\bibfnamefont {H.}~\bibnamefont {Choi}}, \bibinfo {author}
  {\bibfnamefont {T.-J.}\ \bibnamefont {Lu}}, \bibinfo {author} {\bibfnamefont
  {A.~E.}\ \bibnamefont {Dane}}, \bibinfo {author} {\bibfnamefont
  {D.}~\bibnamefont {Englund}}, \ and\ \bibinfo {author} {\bibfnamefont
  {K.~K.}\ \bibnamefont {Berggren}},\ }\href@noop {} {\bibfield  {journal}
  {\bibinfo  {journal} {Nature Nanotechnology}\ }\textbf {\bibinfo {volume}
  {13}},\ \bibinfo {pages} {596} (\bibinfo {year} {2018})}\BibitemShut
  {NoStop}%
\bibitem [{\citenamefont {Heuck}\ \emph {et~al.}(2018)\citenamefont {Heuck},
  \citenamefont {Pant},\ and\ \citenamefont {Englund}}]{Heuck:2018km}%
  \BibitemOpen
  \bibfield  {author} {\bibinfo {author} {\bibfnamefont {M.}~\bibnamefont
  {Heuck}}, \bibinfo {author} {\bibfnamefont {M.}~\bibnamefont {Pant}}, \ and\
  \bibinfo {author} {\bibfnamefont {D.~R.}\ \bibnamefont {Englund}},\
  }\href@noop {} {\bibfield  {journal} {\bibinfo  {journal} {New J. Phys.}\
  }\textbf {\bibinfo {volume} {20}},\ \bibinfo {pages} {063046} (\bibinfo
  {year} {2018})}\BibitemShut {NoStop}%
\bibitem [{\citenamefont {Pant}\ \emph {et~al.}(2017)\citenamefont {Pant},
  \citenamefont {Krovi}, \citenamefont {Englund},\ and\ \citenamefont
  {Guha}}]{Pant:2017ca}%
  \BibitemOpen
  \bibfield  {author} {\bibinfo {author} {\bibfnamefont {M.}~\bibnamefont
  {Pant}}, \bibinfo {author} {\bibfnamefont {H.}~\bibnamefont {Krovi}},
  \bibinfo {author} {\bibfnamefont {D.}~\bibnamefont {Englund}}, \ and\
  \bibinfo {author} {\bibfnamefont {S.}~\bibnamefont {Guha}},\ }\href@noop {}
  {\bibfield  {journal} {\bibinfo  {journal} {Phys. Rev. A}\ }\textbf {\bibinfo
  {volume} {95}},\ \bibinfo {pages} {012304} (\bibinfo {year}
  {2017})}\BibitemShut {NoStop}%
\end{thebibliography}
\end{document}